\documentclass[12pt,a4paper]{article}
\usepackage[colorlinks,citecolor=blue,urlcolor=blue,filecolor=blue,backref=page]{hyperref}
\usepackage{xcolor}
\usepackage{algorithm}
\usepackage{adjustbox}
\usepackage{subfigure}
\usepackage[noend]{algpseudocode}
\usepackage{natbib}
\usepackage{dsfont}
\usepackage{amsfonts}
\usepackage{amssymb}
\usepackage[utf8]{inputenc}
\usepackage{amsmath}
\usepackage{natbib}
\usepackage[paperheight=12in,paperwidth=10in]{geometry}

\newcommand\independent{\protect\mathpalette{\protect\independenT}{\perp}}
\def\independenT#1#2{\mathrel{\rlap{$#1#2$}\mkern2mu{#1#2}}}

\usepackage{array}
\newcolumntype{H}{>{\setbox0=\hbox\bgroup}c<{\egroup}@{}}

\usepackage{multirow}
\usepackage{epigraph,xcolor}

\DeclareMathOperator{\logit}{logit}

\begin{document}
	
	\title{\textbf{\Large Causal Analysis at Extreme Quantiles with Application to London Traffic Flow Data}}
	
	\date{}
	\maketitle
	\author{
		\begin{center}
			\vskip -1cm
	\textbf{Prajamitra Bhuyan\footnotemark[1]$^{\mathsection}$, Kaushik Jana \footnotemark[1]$^{\S}$, Emma J. McCoy$^{\ast, \dag}$}\\
	$^{\mathsection}$Indian Institute of Management, Calcutta, India\\
	$^{\S}$ Ahmedabad University, India\\
	$^{\ast}$London School of Economics and Political Science, United Kingdom\\
	$^{\dag}$The Alan Turing Institute, United Kingdom\\
		\end{center}
		\footnotetext[1]{The authors contribute  equally to this paper.}
	}

\begin{abstract}
Transport engineers employ various interventions to enhance traffic-network performance. Quantifying the impacts of Cycle Superhighways is complicated due to the non-random assignment of such an intervention over the transport network. Treatment effects on asymmetric and heavy-tailed distributions are better reflected at extreme tails rather than at the median.  We propose a novel method to estimate the treatment effect at extreme tails incorporating heavy-tailed features in the outcome distribution. The analysis of London transport data using the proposed method indicates that the extreme traffic flow increased substantially after Cycle Superhighways came into operation.

\end{abstract}


\section{Introduction}
In the last couple of decades, metropolitan areas in both the developed and the developing countries have been affected by increasing traffic congestion mainly due to population explosion. Urban population in developing nations is projected to grow, adding 2.5 billion people by 2050 \citep{UN}. In  addition to the negative impacts on mobility and air quality, previous studies indicate that severe congestion has a negative impact on GDP, the city's economic competitiveness \citep{Cost, Growth}, and road safety, thus imparting socio-economic distress \citep{Retallack}. Intelligent transportation systems can revolutionize traffic-mobility management and offer an integrated approach to infrastructure development \citep{Xian_2021}.  Transport engineers employ network interventions (i.e. treatment) to control high-consequence traffic events. Often such interventions fail to achieve the intended objective, and as a consequence transport networks perform poorly concerning traffic flow, capacity utilisation, safety, and economic and environmental impacts. The underlying reason for such failure could be that the interventions are often made without statistical evidence to guarantee the intended outcomes. In general, the interventions are typically targeted to address specific network problems, and the challenge is that their effect is `confounded' if the treated and control units differ systematically with respect to several characteristics which may also affect the outcome of interest. 

%In the last couple of decades, metropolitan areas in both developed and developing countries, have been affected by increasing traffic congestion due to population explosion.  Urban population in developing nations is projected to continue to grow, adding 2.5 billion people to the world’s cities by 2050 \citep{UN}. In  addition to the negative impacts on mobility and air quality, previous studies indicate that severe congestion has a negative impact on GDP, the city's economic competitiveness \citep{Cost, Growth}, and road safety, thus imparting socio-economic distress \citep{Retallack}. Intelligent transportation systems can revolutionize traffic-mobility management and offer an integrated approach to infrastructure development \citep{Xian_2021}. World-wide transport engineers employ network interventions (i.e. treatment) as a measure to control high-consequence events. But, such decisions are often made without any statistical evidence to guarantee the intended outcomes. Often such interventions fail to achieve the intended objective and in which transport networks perform poorly in relation to traffic flow, capacity utilisation, safety, and economic and environmental impacts. In general, the interventions are typically targeted to address specific network problems, and their effect is `confounded' if the treated and control units differ systematically with respect to several characteristics which may affect the outcome of interest. 

\begin{figure}[htp]	
	\centering
	\includegraphics[scale=0.2]{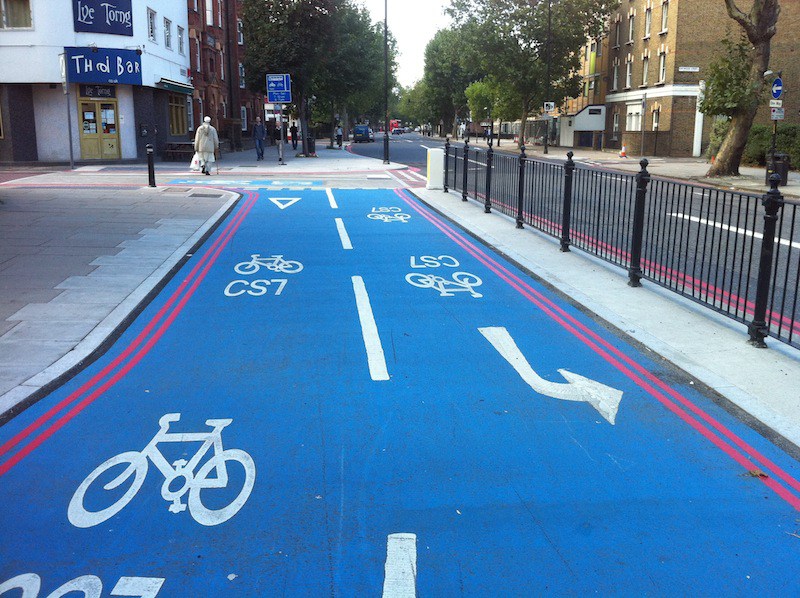}
	\includegraphics[scale=0.2]{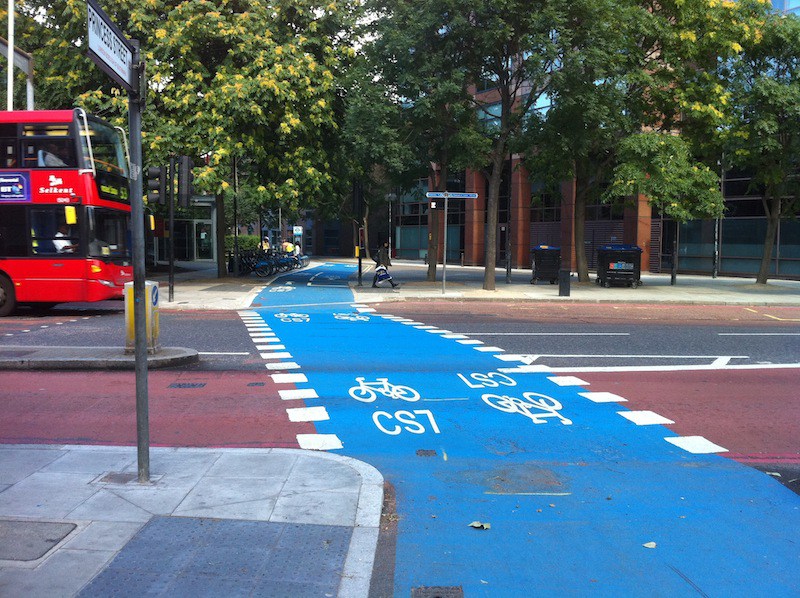}
	
	\includegraphics[scale=0.2]{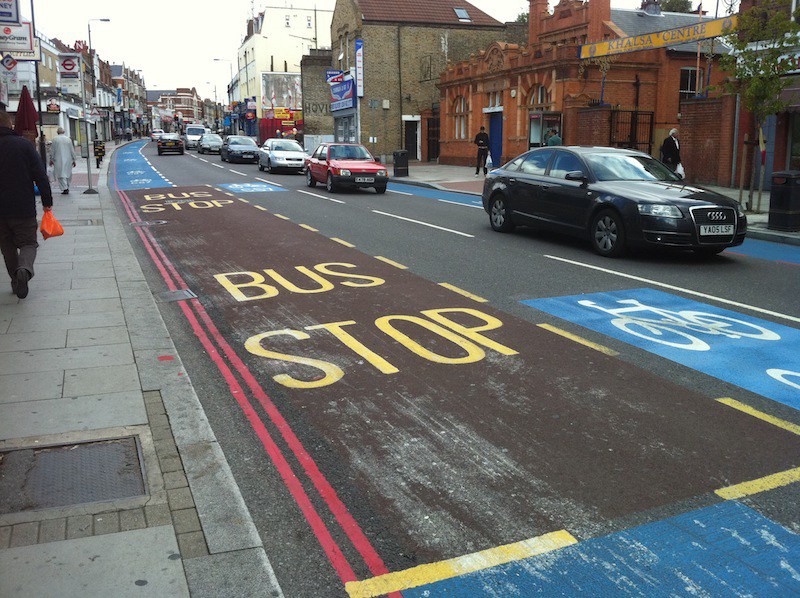}
	\includegraphics[scale=0.2]{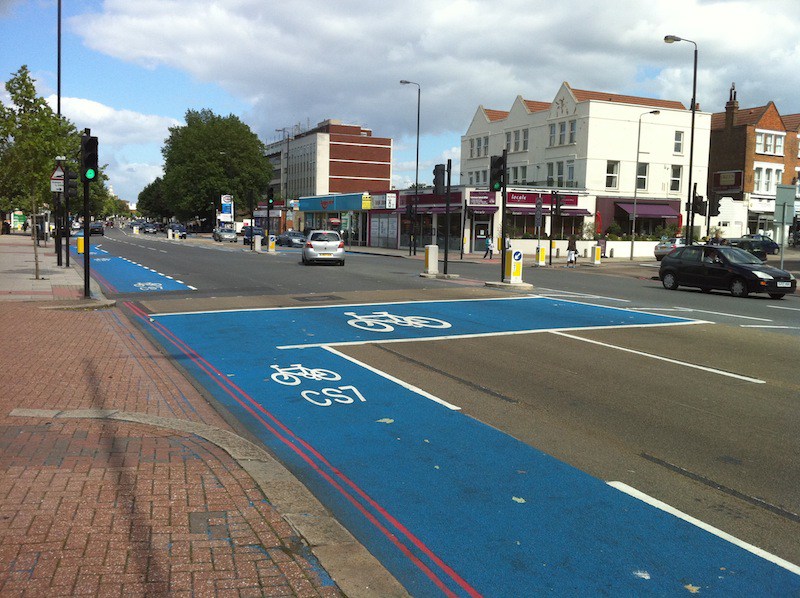}
	\caption{Examples of London Cycle Superhighways}
	\label{CS}
\end{figure}

\begin{figure}[htp]	
	\centering
	\includegraphics[scale=0.4]{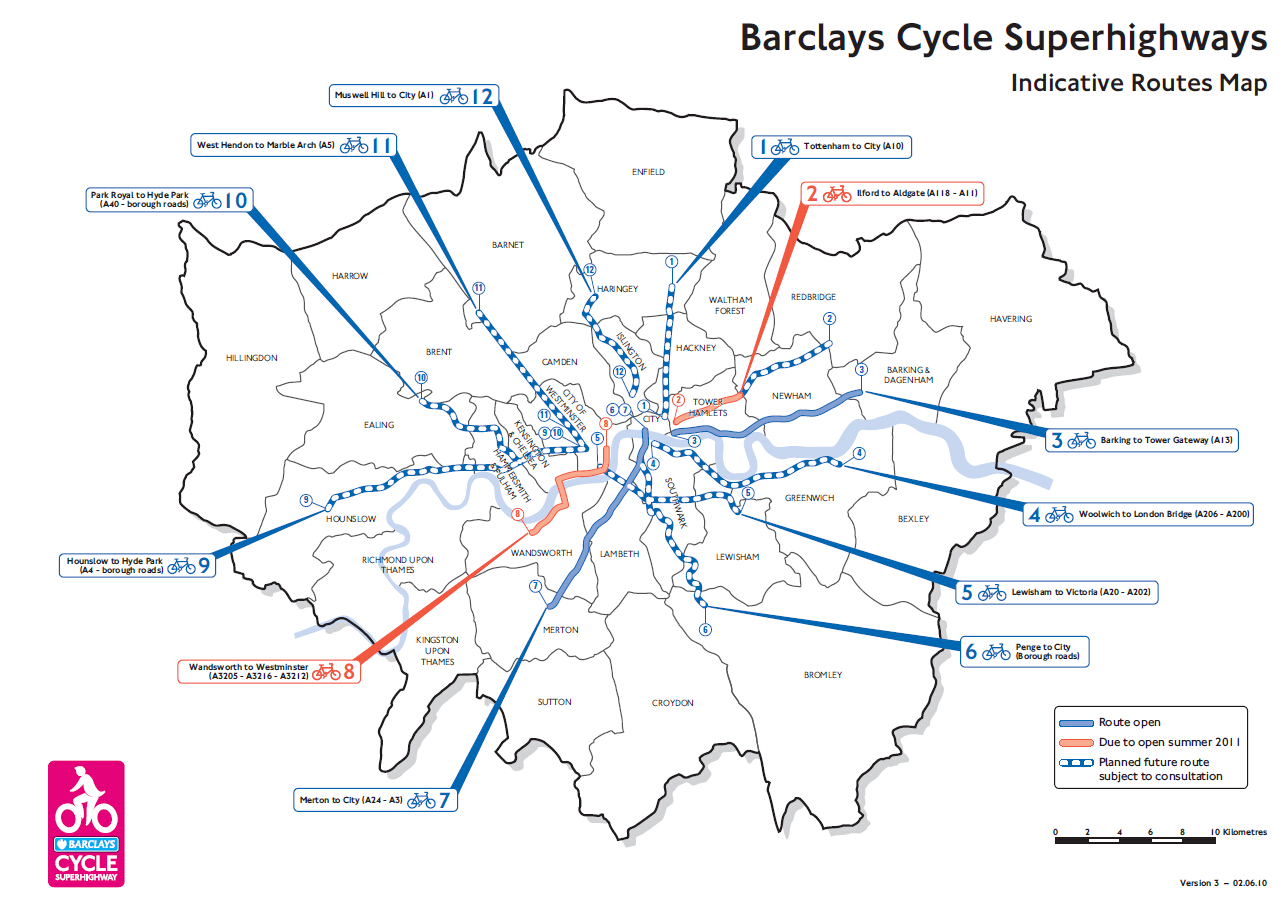}
	\caption{Route map of London Cycle Superhighways}
	\label{Route}
\end{figure}

In the last decade, cycling has been promoted as an affordable, clean and environment friendly sustainable means of transportation arond the world. The bicycle contributes to cleaner air and less congestion that promotes economic growth and reduces inequalities while bolstering the fight against climate change to achieve sustainable development goals \citep{Cycle_day}. A recent study on Traffic Scorecard released in 2021 ranks London as the world’s most congested city out of more than 1,000 global cities \citep{Forbes}. The Mayor of London aims for cycling journeys in London to increase from 2\% of all journeys in 2001 to 5\% by 2026. To promote cycling activity, several policy decisions including Cycle Superhighways (CS), Santander Cycles and Biking Boroughs have already been implemented \citep{Mayor}. The CS are 1.5-meter wide barrier-free cycling paths connecting outer London to central London to provide adequate spatial capacity for existing cyclists and potential future commuters who adopt cycling as a mode of transport (see Figure \ref{CS}). Twelve Cycle Superhighways were announced in 2008 with the aim of enabling faster and safer cycle journeys. As displayed in Figure \ref{Route}, these routes were designed in a clock faced layout to radiate from the city center towards greater London. Currently, only four routes are in operation, namely CS2 (Stratford to Aldgate); CS3 (Barking to Tower Gateway); CS7 (Merton to the City); and CS8 (Wandsworth to Westminster). In the very first year, cycling has increased by $83\%$ along CS3 and $46\%$ along CS7 after their inauguration in 2010.
%In July 2010, the first two pilot routes, CS3 and CS7, were inaugurated. As reported by \citet{TFL11}, in the first year, cycling has increased by $83\%$ along CS3 and $46\%$ along CS7. A new East-West route was introduced to replace CS10, while CS6 and CS12 have been cancelled. 
The effects of CS on heavy congestion are not evaluated in the report by \citet{TFL11} due to the lack of adequate data and methodological framework for analysis. However, some contradictory reports about the effects of CS on road traffic congestion have been published in print and electronic media \citep{Guardian, Evening}. Due to the intricate nature of the transport network, the quantification of the effects of CS on high consequence events like extreme traffic congestion is a complex problem. Moreover, various traffic and socioeconomic characteristics may act as confounders, affecting both the traffic flow and intervention simultaneously, therefore it is important to study the association of those factors with the interventions and outcome of interest. 
%For example, business centres with high employment density with interactively connecting roads are expected to have more congestion compared to the residential areas.

%{\bf The study of causal effect of the treatment on the response in the presence of covariates is falls under the methodology of causal inference and has wide application} in various fields of science, e.g., medical science \citep{Imbens_2015}, transport engineering \citep{Li_2017, Zhang_2021},  and public policy \citep{Gangl_2013, David_2010}. 
The need to develop statistical methods to study causal quantities is important due to their wide application and impact in various fields of science, e.g., medical science \citep{Imbens_2015}, transport engineering \citep{Li_2017, Zhang_2021},  climate science \citep{Hannart2018,Naveau2020review} and public policy \citep{Gangl_2013, David_2010}. 
Most commonly, the literature on causal inference is focused on population means of potential outcomes. In many real scenarios, asymmetric and heavy-tailed distributions are frequently encountered and treatment effects are often better summarized in  tail quantiles rather than in averages. There is a growing literature that focuses directly on the estimation of casual effects on quantiles. \citet{Firpo_2017} proposed estimating the quantiles by minimizing an inverse probability weighted check loss function, which achieves non-parametric consistency by means of a propensity score estimated as a logistic power series with degree increases asymptotically. \citet{Melly_2006}, and \citet{Frolich_2013, victor2013} considered estimation of the quantiles under a linear parametric model for the distribution function, and quantile functions, respectively. \citet{zhang2012} developed several methods for estimation of causal effects on quantiles which are analogous to the methods employed for the average causal effect.
\citet{Diaz_2017} devised a doubly robust estimator based on a semiparametric approach using a targeted maximum likelihood estimator for the quantiles. In this context, \citet{Xu_2018} introduced a Bayesian nonparametric approach through Bayesian additive regression trees to model the propensity score and a Dirichlet process mixture of normal models to construct the distribution of potential outcomes. 

Estimation of extreme quantiles (or equivalently, so-called return levels) of univariate and conditional distributions is an important problem in various application areas, including meteorology, hydrology, climatology and finance \citep{Embrechts1997}. Extreme value theory provides tools to estimate the probability of extreme events based on the assumption that the underlying distribution of the normalized random variable resides in the domain of attraction of extreme value distribution \citep{fisher_tippett_1928}. Thus extreme quantiles can be predicted by estimating the parameters of the corresponding extreme value distribution and the normalizing constants. An alternative approach is to characterise the distribution with its tail index and utilise the same for estimating an extreme quantile. For a detailed review on different models and methods, the readers are referred to the classical books on extreme value inference by \cite{Embrechts1997} and \cite{Coles_2001}, and the review paper by \cite{Gomes2015}. In the context of causal inference, 
\cite{YichongZhang} considered the estimator proposed by \citet{Firpo_2017} and developed asymptotic theory for the causal estimator for intermediate and extreme quantiles. Very little work exists in the literature relating causality and the occurrence of extreme events. Recently \cite{gnecco2019causal} and \cite{Engelke2020causal} combined causal inference with extreme value theory to characterise causal tail dependence of two random variables in the context of a directed acyclic graphical model. \cite{gissibl2018,gissibl2017tail} studied the causal dependence structure through max-linear models on acyclic graphs.  For a hydrological application, \cite{mhalla2019} studied the causal relationship between two extreme random variables for modelling daily measurements of river discharges corresponding to riverflow-connected stations. This modelling approach uses a copula structure to estimate the dependence between two random variables with the restriction that there are no confounders (common set of covariates affecting treatment and response simultaneously) in the model. \citet{Deuber2022} developed a propensity score based estimator of  extremal quantile treatment effect. This method relies on asymptotic tail approximations through the Hill estimator for the extreme value indices of potential outcome distributions. Recently, there has been a focus on studying the causal links between  the climate system and climate change (attribution) and this causal link has been used to design event attribution; see \cite{Hannart2018} and \cite {Ribes2020} for more details. 
%\cite{Naveau2020review} provides a comprehensive overview of the literature on extreme event attribution in climate science.

\subsection{London transport-network data}\label{Data}
The current study is motivated by the problem of estimation of the causal effect of London Cycle Superhighways on extreme traffic congestion based on a data set over the period 2007-2014. The data consists of 75 treated zones and 375 control zones which were selected through stratified random sampling along the 40 km long main corridors radiating from central London to outer London. In observational data, the effect of CS is confounded due to various factors related to traffic dynamics, road characteristics, and socio-demographic conditions. Traffic data on both the major and minor road network are routinely collected by the Department for Transport,  Government of the UK. Additional data on traffic flow and speed are collected from the London Atmospheric Emissions Inventory. It has been observed that traffic congestion is associated with bus-stop density and road network density \citep{Bus}. An association between traffic congestion and socio-demographic characteristics, such as employment and land-use patterns, has also been indicated in previous studies \citep{Land1, Employ, Land2, Bhuyan2021}. A detailed description of the above data sets can be found in \citet{Li_2017}. To incorporate these socio-demographic effects, we obtained relevant data on population and employment density, as well as the information of land-use patterns from the Office for National Statistics. The data that are available from the aforementioned sources and the rationale to construct the response and covariates are described in Section \ref{Application}.

%We are interested in estimation of the causal effect of London Cycling Superhighways on extreme traffic congestion based on a data set collected over the period 2007-2014. In this study, 75 treated zones and 375 control zones were selected using stratified random sampling along the 40 km long main corridors radiating from central London to outer London. In observational data, the effect of CS is confounded due to various factors related to traffic dynamics, road characteristics, and socio-demographic conditions. Traffic data on both the major and minor road network are routinely collected by the UK Government’s Department for Transport. Also, additional information on traffic flow and speed are collected from the London Atmospheric Emissions Inventory (LAEI). It is observed that traffic congestion is associated with bus-stop density and road network density. An association between traffic congestion and socio-demographic characteristics, such as employment and land-use patterns, has also been indicated in previous studies \citep{Land1, Employ, Land2, Bhuyan2021}. See \citet{Li_2017} for the detailed data description. To incorporate these effects, we obtained relevant data on population and employment density, as well as the information of land-use patterns from the Office for National Statistics. The data that are available from the aforementioned sources and the rationale to construct the response and covariates are described in Section \ref{Application}.

Recent literature on transport engineering emphasized the importance of developing inferential tools to predict extreme consequences of interventions or shocks to the system, enabling one to identify the most effective available interventions to mitigate the consequences of such events \citep{Zheng_2019, Xu_2017, FarahaCarlos_2017}. As mentioned above, the motivating data on traffic system arises from a study of various interventions targeted at traffic conditions in the city of London. A vital task for the policy makers is to quantify the causal effects of these interventions on the extreme traffic flow and investigate its associated consequnces. The outcome variable (traffic flow) under consideration exhibits heavy tails as there are small but non-trivial number of locations with very high traffic volume. Heavy-tailed distributions are often characterized by very large variance. For such cases, standard error of the existing estimators of the causal quantities are also very high and subsequently precludes statistical significance at most plausible sample sizes \citep{Diaz_2017}. In this work, we address this issue and propose a method of estimation of the causal effect at the extreme tails by combining tools of causal inference and extreme value theory.
%\textcolor{red}{The small sample performance of the  proposed method and its gain over competing methods are demonstrated through extensive simulation studies.} 
The method is applied to analyse the motivating data from the transport engineering application outlined above.

The outline of the article is as follows, in Section \ref{Method} we describe the existing models and methods used in the traditional potential outcome framework. In Section \ref{Prop_method}, we propose a new method to estimate the causal effects at extreme quantiles. Simulation studies are performed to assess the effectiveness of the proposed methods and the results are summarized in Section \ref{Sim}. In Section \ref{Application}, we discuss the results obtained by analysing the London Cycle Superhighways data. We summarise the key findings and conclude with a discussion of future research in Section \ref{Discussion}.

\section{Potential outcome framework}\label{Method}
In usual causal inference problems, the data available for estimation of the causal effects are realisations of a random vector, $Z_i = (Y_i,D_i,\boldsymbol{X_i})$, $i = 1,\ldots, n$, where for the $i$th unit of observation, $Y_i$ denotes the response, $D_i$ the treatment (intervention or exposure) received, and $\boldsymbol{X_i}$ a vector of confounders or covariates. Let the triplet $(Y, D, \boldsymbol{X})$ represents a generic form of $(Y_i, D_i, \boldsymbol{X}_i)$, $i = 1,\ldots, n$. The treatment can be binary, multi-valued or continuous but essentially it is not assigned randomly. Therefore, the simple comparisons of average responses across different treatment groups will not in general reveal a ‘causal’ effect (causation) due to potential confounding. Confounding can be addressed if the vector of covariates $\boldsymbol{X}$ is sufficient to ensure unconfoundedness, or conditional independence of potential outcomes and the treatment. In the context of binary treatment, the conditional independence assumption requires that $(Y(0), Y(1)) \independent D|\boldsymbol{X}$, where $D$ is the indicator function for receiving the treatment, and $Y(1)$ and $Y(0)$ indicate potential outcomes under treated or control status, respectively. That is, given the information contained in the covariate vector $\boldsymbol{X}$, the assigned treatment carries no additional information on the individual’s potential outcomes. This assumption is also referred to as conditional exchangeability or strong ignorability in the literature \citep{Rosenbaum1983}. An additional requirement for valid causal inference is positivity or overlap assumption: 
%conditional on covariates $\boldsymbol{X}$,
the conditional probability of treatment assignment is strictly positive for every combination of values of the covariates, i.e. $0<\mathbb{P}\left[D=1| \boldsymbol{X}=\boldsymbol{x}\right]<1$ for all $\boldsymbol{x}$.
This assumption implies that there exists no strata of $\boldsymbol{X}$ such that treatment assignment is uniquely determined.  To illustrate this point, suppose that for a given value of $\boldsymbol{X}$, say $\boldsymbol{x}^{\ast}$,
$\mathbb{P}\left[D = 0|\boldsymbol{X} = \boldsymbol{x}^{\ast}\right] = 1$.  Then there will be no triplets of observed data $(Y,D,\boldsymbol{X})$ such that $D= 1$ and $\boldsymbol{X}= \boldsymbol{x}^{\ast}$; thus, any estimates of treatment effects, say a conditional average treatment effect 
such as $\mathbb{E}[Y |D = 1,\boldsymbol{X} = \boldsymbol{x}^{\ast}]-\mathbb{E}[Y |D = 0,\boldsymbol{X}=\boldsymbol{x}^{\ast}]$ 
would require extrapolation. The readers are referred to \cite{Erica_2022} for more details. Conventionally, the main interest is in estimating the average treatment effect 
$\mu = \mathbb{E}[Y(1)]-\mathbb{E}[Y(0)]$, 
which measures the difference in average outcomes under treatment and control status.

In practice, distributional features are often summarized using quantiles when asymmetric distribution for the response variable is encountered. In case of a skewed distribution, the median is a more appropriate measure of location than the mean, and statistics based on quantiles could be more meaningful measures of the spread of the distribution than the standard deviation. In many applications, practitioners are interested in the tails of the distribution (say higher than 95th percentiles) of the potential outcomes \citep{zhang2012, Diaz_2017}. In the presence of confounding, a simple comparison
of the treated and untreated individuals in terms of empirical quantiles would not have a causal interpretation. In such situation, the causal effects of treatment on the target $\tau_0$-th $(0<\tau_0<1)$ quantile is defined as 
\begin{equation}\label{target}
	\eta_{\tau_0}=q_{Y(1)}(\tau_0)-q_{Y(0)}(\tau_0),
\end{equation} 
where $q_{Y(t)}$ denotes the quantile associated with the distribution of potential outcome $Y(t)$ for $t=0,1$ \citep{zhang2012}. In many contexts, it is also of interest to understand the quantile effect of applying versus withholding the intervention only for the treated units (rather than for all the units in the population). This causal quantity is called the quantile treatment effect on the treated and is defined as \begin{equation}\label{target_t}
	\zeta_{\tau_0}=q_{Y(1)|D=1}(\tau_0)-q_{Y(0)|D=1}(\tau_0),
\end{equation}
where $q_{Y(t)|D=1}$ denotes the quantile associated with the distribution of the potential outcome $Y(t)$ given $D=1$ for $t=0,1$ \citep{zhang2012}.
When there exists no heterogeneity of the treatment effect induced by the covariates (i.e., no interactions effects of the treatment and the confounders), the quantile treatment effect  on the population and the quantile treatment effect on the treated are identical provided the effects are additive. However, when treatment heterogeneity exists, these two treatment effects differ \citep{Erica}.

Causal effects on quantiles can be estimated using quantile regression with appropriate adjustment for confounders. However, it should be noted that the quantile estimates from a quantile regression model are conditional on all the covariates in the model. When one is interested in estimating the marginal quantile at a given probability level (without conditioning on covariates), it has to be determined by all conditional quantiles. \cite{zhang2012} studied the problem of estimating the causal quantities, defined in (\ref{target}) and (\ref{target_t}), under parametric assumption on the distribution of the outcome variable.
%Let $(Y, D, \boldsymbol{X})$ represent a generic form of the triplet $(Y_i, D_i, \boldsymbol{X}_i)$, and similarly, let $Y(t)$ denote the potential outcome of a generic unit for $i=1, \ldots, n$, $t=0$ or 1. 
In light of the conditional exchangeability assumption, the distribution of the potential outcome $Y(t)$, for $t=0,1$, (i.e. counterfractual distribution) is given by
\begin{eqnarray}\label{marginal_y}
	F_t(y)
	&=& \mathbb{E}\left[P(Y(t)\leq y|\boldsymbol{X})\right]\nonumber\\
	&=& \mathbb{E}\left[P(Y(t) \le y |D=t, \boldsymbol{X}) \right]\nonumber\\
	&=& \mathbb{E}\left[P( Y \le y |D=t, \boldsymbol{X}) \right]\nonumber\\
	&=&\int G_{t,x}(y)d H(\boldsymbol{x}),
\end{eqnarray} 
where $G_{t,x}(y)=P(Y\leq y | D=t, \boldsymbol{X}=\boldsymbol{x})$ for $y\in \mathbb{R}$ , and $H$ denotes the marginal distribution of $\boldsymbol{X}$ for $t=0, 1$.
Similarly, the distribution of the  potential outcome $Y(t)$ conditional on $D=1$, is given by
\begin{equation}\label{marginal_y_t}
	F_{t|1}(y)=E\{P(Y(t)\leq y|\boldsymbol{X})|D=1\}=\int G_{t,x}(y)d H_{1}(\boldsymbol{x}),
\end{equation}  
where $H_{1}$ denotes the marginal distribution of $\boldsymbol{X}$ given $D=1$ for $t=0,1$.

\cite{zhang2012} estimated the counterfactual distribution, given in (\ref{marginal_y}), by 
\begin{equation}\label{marginal_est}
	\hat F_t(y)=n^{-1}\sum_{i=1}^n\hat{G}_{t,\boldsymbol{X_i}}(y),
\end{equation} 
where $\hat G$ is an estimate of $G$ under the assumption that a normal linear model holds after a Box–Cox transformation. Finally the target $\tau_0$-th quantile $q_{Y(t)}(\tau_0)$ is estimated by inverting the estimated unconditional distribution function $\hat F_t(y)$ and substituting it into the expression given in (\ref{target}) to obtain the estimate of the quantile treatment effect for the population. This is known as outcome regression (OR) based estimate. Similarly, one can estimate $F_{t|1}(\cdot)$, given in (\ref{marginal_y_t}), by
\begin{equation}\label{marginal_est_terat}
	\hat F_{t|1}(y)=\frac{1}{^{\sum_{i=1}^{n}D_i}}\sum_{i=1}^{n}\hat{G}_{t, \boldsymbol{X_i}}(y)D_i.
\end{equation} 
One can then invert $\hat F_{t|1}(y)$ to obtain estimates of $q_{Y(t)|D=1}(\tau_0)$ and the quantile treatment effect for the treated $\zeta_{\tau_0}$ defined in (\ref{target_t}). 
An alternative set of estimators of the counterfactual distributions proposed by \cite{zhang2012} are based on inverse propensity weighting (IPW) and matching techniques.
%An alternative set of estimators of the counterfactual distributions is based on inverse propensity weighting (IPW) and matching techniques \citep{zhang2012}.
\citet{Firpo_2017} proposed quantile regression approach for estimation of counterfactual quantiles using inverse propensity score as weights.

\section{Proposed methodology}\label{Prop_method}
Note that the conditional distribution of $Y(t)$ given $\boldsymbol{X}=\boldsymbol{x}$ can be obtained from the conditional quantile function $Q_{Y|t, \boldsymbol{x}}(\tau)=G_{t, \boldsymbol{x}}^{-1}(\tau)$ of the outcome $Y$ for all $\tau \in (0,1)$ \citep{victor2013}. Let $U\sim Unif(0,1)$. Then we can write
\begin{eqnarray}\label{pop-dist}
	G_{t,\boldsymbol{x}}(y)&=& P\left[U \le G_{t,\boldsymbol{x}}(y) \right] \nonumber\\
	&=& P\left[Q_{Y|t, x}(U) \le y \right] \nonumber\\
	&=& E\left[\mathbb{I}\left[Q_{Y|t, x}(U)\le y\right]\right]\nonumber\\
	&=&\int_{0}^1  \mathbb{I}\left[Q_{Y| t, \boldsymbol{x}}(\tau)\leq y\right]d\tau,
\end{eqnarray}
where $\mathbb{I}[\cdot]$ is an indicator function. When the context suggests the presence of a heavy-tailed nature in the data generating process and one is interested in the extreme tails of the distribution $F_t$ (i.e. probability levels are close to 0 or 1), it is more appropriate to model $G_{t,\boldsymbol{x}}$ with some heavy-tailed components.  To model the bulk and the extreme component of $G_{t,\boldsymbol{x}}(y)$ efficiently, we choose a suitable transition probability level $\tau_u\in (0, 1)$ and decompose (\ref{pop-dist}) as:
\begin{eqnarray}\label{decomposition}
	G_{t,\boldsymbol{x}}(y)=\int_{0}^{\tau_u}  \mathbb{I}\left[Q_{Y|t,\boldsymbol{x}}(\tau)\leq y\right]d\tau+\int_{\tau_u}^{1} \mathbb{I}\left[Q_{Y|t,\boldsymbol{x}}(\tau)\leq y\right]d\tau.
\end{eqnarray}
The transition point $\tau_u$ is assumed to represent the probability level of quantile above which the distribution (second part of \eqref{decomposition}) exhibits heavy-tailed nature. %Identification of this point is a crucial issue. 
%Note that the transition point needs to be high enough such that the second summands of \eqref{decomposition} exhibit heavy-tailed features. 
At the same time a very high probability level as a choice of $\tau_u$, may prevent enough number of observations being generated from the extreme part of the distribution. In practice, $\tau_u$ needs to be estimated from the data by balancing the above two issues. The next section provides a detailed outline of a data driven and efficient error control approach to determine the probability level $\tau_u$. 
Note that, any estimation of the unknown population quantile process involves the evaluation of the quantile process at finite grid points. Thus the integral in \eqref{decomposition} needs to be evaluated at discrete points. A common requirement to reliably estimate the quantile process is that the mesh width of grids should converge to 0 with a rate $1/\sqrt{n}$ or faster \citep{victor2020}.

We propose an estimate of $G_{t,\boldsymbol{x}}(\cdot)$ as 
\begin{eqnarray}\label{sample-decomposition}
	\hat{G}_{t,\boldsymbol{x}}(y)
	=\sum_{j=1}^{u} (\tau_j-\tau_{(j-1)})\mathbb{I}\left[\hat{Q}_{Y|t,\boldsymbol{x} }(\tau_{j})\leq y\right]+\sum_{j=u+1}^J (\tau_j-\tau_{(j-1)})\mathbb{I}\left[\tilde{Q}_{Y|t,\boldsymbol{x}}(\tau_{j})\leq y\right],
\end{eqnarray}
where the integer $u$ is determined from $\tau_u$, and $\hat{Q}_{Y|t,\boldsymbol{x}}(\tau_{j})$ and $\tilde{Q}_{Y|t,\boldsymbol{x}}(\tau_{j})$  are estimates of $Q_{Y|t,\boldsymbol{x}}(\tau_{j})$ based on simultaneous quantile regression and parametric extreme value modelling, respectively, such that the issue of quantile crossing (within each of the summands) does not arise. The details of the above estimation methods are discussed in Subsections \ref{sec:transition}-\ref{bulk}.
The estimated conditional distribution function is then plugged into \eqref{marginal_est} to obtain the estimate of counterfactual distributions $\hat F_t(y)$ for $t=0, 1$. Then $\hat F_t(y)$ is inverted to find the quantile $\hat q_{Y(t)}(\tau_0)=\inf\{z: \hat{F}_t(z)\geq \tau_0\}$ for $t=0,1$, and finally the estimate of quantile causal effect $\eta_{\tau_0}$ for the population is given by
\begin{equation}\label{target_estimate}
	\hat \eta_{\tau_0}=\hat q_{Y(1)}({\tau_0})-\hat q_{Y(0)}({\tau_0})\nonumber.
\end{equation} 
Similarly, we obtain $\hat q_{Y(t)|D=1}({\tau_0})=\inf\{z: \hat{F}_{t|1}(z|D=1)\geq {\tau_0}\}$ from (\ref{marginal_est_terat}), and the estimate of $\zeta_{{\tau_0}}$ is given by 
\begin{equation}\label{target_estimate_t}
	\hat \zeta_{{\tau_0}} = \hat q_{Y(1)|D=1}({\tau_0})-\hat q_{Y(0)|D=1}({\tau_0}) \nonumber.
\end{equation} 

\subsection{Selection of transition point}\label{sec:transition} 
The value of $\tau_u$th quantile in \eqref{decomposition} determines the transition point of the outcome distribution, below which the quantile regression model is fitted to the bulk part of the distribution, and above which a heavy-tailed distribution is fitted. Let $Y$ be a random variable with distribution function $\Psi$ with right end point $y_{\Psi}$. Then by the Pickands-Balkema-de Haan Theorem \citep{BalkemadeHaan1974GPD, Pickands_1975}, the conditional distribution of $Y-U$ above an excess $U<y_{\Psi}$ is well approximated by the generalized Pareto distribution (GPD) for large $U$ under some general regularity conditions. That is, 
%Let us consider the conditional distribution of $Y-U$ above an excess $U<y_{\Psi}$. Then under some general regularity conditions and by Pickands-Balkema-de Haan Theorem \citep{BalkemadeHaan1974GPD, Pickands_1975}, 
$$\lim_{{y\uparrow y_{\Psi}}} P\left[Y-U\le y|Y>U\right]\rightarrow H(y),$$
and $H$ is the distribution function of GPD, given as:
%\begin{equation}\label{gpd0} H(y)=\begin{cases} 1-\left[1+\xi \left( \frac{y}{\sigma}\right)\right]^{-1/\xi},\;\; y \geq 0, \xi\geq 0,\\1-\exp\left[\frac{y}{\sigma}\right],\;\;\;\; 0<y<-\sigma/\xi, \xi<0.\end{cases}\end{equation}
\begin{equation}\label{gpd0}
	H(y)=\begin{cases} 1-\left[1+\xi \left( \frac{y}{\sigma}\right)\right]^{-1/\xi},\;\; \mbox{if}\;\;\xi\neq 0,\\
		1-\exp\left[\frac{y}{\sigma}\right],
		\;\;\;\; \;\;\;\;\;\;\;\;\;\;\mbox{if} \;\;\;\xi=0 \nonumber,
	\end{cases}
\end{equation}
where 
$y\geq 0 \;\mbox{if} \;\;\xi\geq 0$, and
$0\leq y\leq -\sigma/\xi\;  \;\mbox{if}\;\; \xi<0.$
Here $\sigma$ and $\xi$ are the scale and shape parameters, respectively. 
The value of the shape parameter $\xi=0$, interpreted as $\xi\rightarrow 0$, leads $H$ to be the exponential distribution with mean $\sigma$. The negative values of $\xi$ imply the right tail of $H$ is short and light, and it has a finite right end point. In particular, for $\xi=-1$, $H$ is a uniform distribution over $[0, \sigma]$.  When $\xi>0$, the higher the value of $\xi$ the heavier the tail of the distributions. Choice of the threshold and fitting of a GPD to the exceedences of the threshold is a widely studied problem in the extreme value analysis literature. Too low and too high value of the threshold should generally be avoided \citep{Coles_2001}. A very high threshold leaves fewer exceedances to fit the GPD model and would yield estimates with high variance, and a too low threshold leads to an inadequate fit to the GPD model and may produce biased estimates. In the light of the decomposition of the estimate $\hat{G}_{t,x}$, provided in \eqref{sample-decomposition}, the threshold requires to be a suitable quantile of the conditional distribution of the outcome. Many threshold selection procedures are available in the literature; see  \cite{Scarrott_2012} for a review. 

There are broadly four categories of approaches for threshold selection. The first, and most common approach is based on a graphical diagnosis method \citep{Davison1990, drees2000,Coles_2001}. The second approach involves methods that minimize the asymptotic mean-squared error of the estimators of the GPD parameters or of the extreme quantiles, under particular assumptions on the upper tail of the GPD \citep{hall_1990, beirlant_2004,Langousis2016}. 
The third category of methods are based on the goodness-of-fit (gof) test of the GPD \citep{Davison1990, bader2018}, where the threshold is selected at a level to ensure that the GPD provides an adequate fit to the exceedances.  The fourth method is automatic threshold selection method of \cite{bader2018} for choosing a suitable threshold. Most of the above methods can incorporate additional covariate information in different levels of the modelling. See \cite{Davison1990} and  \cite{Beirlant2005} for an extensive review of covariate dependent GPD modelling.
%We use the quantile regression method \cite{koenker2005} and use covariate dependent conditional quantile as the threshold. 

%{\bf Two-step approach for selection of $\tau_u$:}
\subsubsection{Two-step approach for selection of $\tau_u$}
To select the transition point $\tau_u$ in \eqref{pop-dist}, we develop a two-step approach. In the first step we use covariates to generate multiple competing and ordered thresholds through constraint quantile regression (discussed in the following Section~3.2). And in the second step, we adopt the automatic threshold selection method of \citet{bader2018} controlling false discovery rate associated with multiple goodness-of-fit tests. Specifically, in the first step, for a given number $l$, we consider $l$ competing thresholds as  conditional quantiles obtained by quantile regression \citep{koenker2005} and denote the fitted quantiles as $U(\tau|{\boldsymbol{w}})={\boldsymbol{w}}^T\hat{\boldsymbol{\beta}}(\tau)$, where $\boldsymbol{v}=(t, \boldsymbol{x})$ and $\boldsymbol{w}=(1, \boldsymbol{v})^T$ is a covariate vector of dimension $(m+1)$ including 1, and $\hat{\boldsymbol \beta}(\tau)=(\hat\beta_0(\tau),\ldots,\hat\beta_m(\tau))^T$ are the estimates of the regression parameter vector at  $\tau=\tau_1,\ldots, \tau_l$. Therefore, for a given set of $l$ probability levels, the method produces thresholds $U(\tau_1|{\boldsymbol{w}})\leq\ldots\leq U(\tau_l|{\boldsymbol{w}})$. 
%In the second step, we adopt the approach of \cite{bader2018} for automatically selecting a threshold from the aforementioned competing set in the presence of covariates. 
We then use each of the thresholds, and fit a GPD to the exceedences above the threshold. A different GPD is obtained for each of the covariate adjusted threshold. We use maximum likelihood method to estimate the unknown GPD parameters. Let us assume that there are $n_k$ exceedences corresponding to the $k$th threshold, for $k=1,\ldots, l$. Then the sequence of null hypotheses to be tested are $H^{(k)}_0$: the distribution of $n_k$ observations above $U(\tau_k | \boldsymbol{w})$ follows the GPD.

\cite{bader2018} adopted and improved the ForwardStop rule of \cite{GSell2015} for the sequential testing (ordered) of null hypotheses $H_0^{(1)},\ldots,H_0^{(l)}$ when the tests are dependent.
The method uses the $p$-values corresponding to the tests of $l$ hypotheses as $p_1,\ldots,p_l\in[0,1]$. For testing multiple hypothesis with an ordered nature, \cite{GSell2015} proposed transforming the sequence of $p$-values to a monotone sequence and then applying the original method of \cite{Benjamini1995} on the transformed sequence. The rejection rule is constructed by returning a cutoff $\hat{k}$ such that $H_0^{(1)},...,H_0^{(\hat {k})}$ are rejected. They prescribe that if no $\hat{k}\in {1,...,l}$ exists, then no rejection is made. ForwardStop is given by 
\begin{equation}
	\hat{k}=\max \left\{k \in \{1,\ldots l\}: -\frac{1}{k}\sum_{i=1}^k\log(1-p_i)\leq \lambda \right\},\nonumber
\end{equation}
where $\lambda$ is a pre-specified level. In this sequential testing, stopping  at $k$ implies that the goodness-of-fit test of the GPD to the exceedances at the first $k$ thresholds ${U(\tau_l|\boldsymbol{w}),...,U(\tau_k|\boldsymbol{w})}$ are rejected. In other words, the set of first $k$ null hypotheses $H_0^{(1)},...,H_0^{({k})}$ are rejected. Let us denote $U(\tau_u|\boldsymbol{w})$ as the GPD threshold corresponding to the selected probability level $\tau_u=\tau_{\hat{k}}$.

\subsection{Estimation of quantiles in the bulk part}\label{bulk} 
After the selection of the transition point between the bulk and the extreme part of the distribution, we need to estimate the conditional quantiles involved in both the summands of \eqref{decomposition}. The first part requires estimation of multiple conditional quantiles at different probability levels. One can estimate these quantiles separately, but this approach has two major drawbacks. Firstly, it could be possible that multiple estimated quantiles cross each other even if their population versions never cross. For example, it may be possible that the estimated 90th percentile has a larger value than the estimated 95th percentile. This results in an invalid estimate of the corresponding distribution function. Secondly, it fails to incorporate common characteristic among the consecutive quantiles \citep{zhu-csda-2008}. Simultaneous estimation borrows information from all the neighbouring quantiles that may improve the efficiency of individual quantile estimates \citep{Liu_2010}.  In their work, \cite{Sabine_2013} used splines to model the quantile regression functions simultaneously but, the drawback with the method is that there is no clear theoretical justification which guarantees that there will be no crossings. 

Alternatively, the method of \cite{Bondell_2010} ensures non-crossing for simultaneous quantile estimation (first part of \eqref{sample-decomposition}). Let $\boldsymbol{C} \subset \mathbb{R}^m$, be a closed convex polytope, represented as the convex hull of $n$ points in $m$-dimensions. Recall that $\boldsymbol{v}=(D, \boldsymbol{x})$ and $\boldsymbol{w}=(1, \boldsymbol{v})^T$, and we are interested in ensuring that the quantile curves do not cross for all values of the covariate $\boldsymbol{v}\in \boldsymbol{C}$. For estimating multiple quantiles at probability levels $\tau_1<,\ldots,<\tau_u$, \cite{Bondell_2010} devised a simple and computationally efficient procedure using a version of constrained optimization that gureentees non-crossing of the estimated quantiles for all values of the covariate $\boldsymbol{v}\in \boldsymbol{C}$. This goal is achieved by imposing appropriate dominance in every consecutive pair of quantile regression planes. Under this setup, the parameters associated with the quantile function are estimated by solving the following optimization problem:
\begin{eqnarray}\label{constrained1}
	&&%\widehat {\boldsymbol \beta}=\underset{\boldsymbol \beta}{\arg\min}
	\hat {\boldsymbol \beta}=\min_{\boldsymbol \beta}
	\sum_{j=1}^u\sum_{i=1}^n
	\rho_{\tau_j}\left(Y_i-\boldsymbol{w}^T_i\boldsymbol \beta(\tau_j)\right),\nonumber\\
	&&\mbox{subject to}\; \boldsymbol{w}^T\boldsymbol \beta(\tau_{j+1})\geq \boldsymbol{w}^T\boldsymbol \beta(\tau_{j})\\
	&&\mbox{for}\quad j=1,\ldots,u-1,\quad \forall \quad \boldsymbol{v}\in \boldsymbol{C},\nonumber
\end{eqnarray}
where $\rho_{\tau_{j}}(w)=\tau_{j}(1-\mathbb{I}(w<0))$ is the check function. Note that, the estimated conditional quantile of $Y$ given $D=t$ and $\boldsymbol{X}=\boldsymbol{x}$ in the first summand of equation \eqref{sample-decomposition} is given by
$\hat{Q}_{Y|t, \boldsymbol{x}}(\tau)=\hat \beta_0(\tau)+t\hat \beta_{1}(\tau)+\boldsymbol{x}^T\hat {\boldsymbol{\beta}}^*(\tau)$, where $\hat{\boldsymbol{\beta}}^*(\tau)$ is the $(m-1)$ dimensional vector of estimated coefficients corresponding to the covariate vector $\boldsymbol{x}$, for $t=0,1$, $\tau=\tau_{1},\ldots, \tau_{u}$.

The proposed method involves estimation of the quantile regression at several probability levels, and it also involves a fixed number of  covariates. For quantile regression with constraints and with the assumption that the covariate space is bounded by the convex polytope, the computation becomes a linear programming problem with $u-1$ constraints \citep{Bondell_2010}. \citet{karmakar1984} presented a polynomial-time algorithm for linear programming that requires $O(m^{3.5}L)$ arithmetic operations on $O(L)$ bit numbers, where $L$ is the number of bits in the input expressed as a function of $u$ and number of covariates $m$.

\subsection{Estimation of quantiles in the extreme part}

Note that the choice of transition point or threshold $U(\tau_u| \boldsymbol{w})$ provided in Subsection \ref{sec:transition} guarantees the suitability of GPD to model the exceedences $Y-U({\tau_u}|\boldsymbol{w})$, 
where 
$U(\tau_u| \boldsymbol{w})=\hat\beta_{0}(\tau_{u})+ t\hat{\beta}_{1}(\tau_{u})+\boldsymbol{w}^{T}\boldsymbol{\hat{\beta}^*}(\tau_{u})$ for $t=0,1$, and $\boldsymbol{\hat{\beta}}^*(\tau_u)$ is the estimated regression coefficient obtained from Section \ref{bulk}. 
Then the estimated conditional quantile of $Y$ given $D=t$, and $\boldsymbol{X}=\boldsymbol{x}$ at probability level $\tau=\tau_{u+1},\ldots, \tau_{J}$, in the second summand of \eqref{sample-decomposition} is given by the approximation (see \cite{Coles_2001}, Section 4.3.3) :
\begin{equation}\label{gpd-est}
	\tilde Q_{Y|t, \boldsymbol{w}}(\tau)\approx \begin{cases}U({\tau_u}|\boldsymbol{w})+\frac{\hat\sigma}{\hat\xi}\left[\left(\frac{\hat\zeta_{U}}{1-\tau}\right)^{\hat\xi}-1\right],& \text{if} \; \hat \xi\neq 0,\\U({\tau_u}|\boldsymbol{w})+\hat\sigma \log\left(\frac{\hat\zeta_{U}}{1-\tau}\right), & \text{if } \hat \xi= 0, 
	\end{cases}
\end{equation} 
where $\hat{\zeta}_{U}=\frac{n_{U}}{n}$, $n_{U}$ is the number of exceedences over the threshold $U({\tau_u}|\boldsymbol{w})$ for the given sample of size $n$, and $\hat\sigma$ and $\hat\xi$ are the maximum likelihood estimates of $\sigma$ and $\xi$, respectively.  Here, the treatment and other covariates influence the aforementioned estimate of the quantiles only through the threshold $U({\tau_u}|\boldsymbol{w})$.

Note that, under the assumption of a heavy-tailed response distribution (with shape parameter $\xi>0$), the above estimated quantile $\tilde Q_{Y|t, \boldsymbol{w}}(\tau)$ in \eqref{gpd-est} is an increasing function of $\tau$. This fact ensures no crossings for the fitted quantiles  at extreme tails. Also, each of the quantile estimate $\tilde Q_{Y|t, \boldsymbol{w}}(\tau)$ is larger than the threshold $U(\tau_u| \boldsymbol{w})$ which is the highest quantile of the bulk part. 
%$\hat{Q}_{Y(t)}(\tau_u|\boldsymbol{X}=\boldsymbol{x})$. 
This fact guarantees that there is no crossing at the transition point between the extreme and the bulk part.

\subsection{Bootstrap algorithm}\label{Boot}
\citet{zhang2012} used the standard full sample bootstrap method to compute the standard error and confidence interval of the proposed estimators (OR, IPW, Matching, etc.) for ease of implementation, avoiding the difficulty of deriving asymptotic variance. However, the validity of the bootstrap method has not been studied. \citet{Zhang_2021} studied the estimator proposed by \citet{Firpo_2017}, and showed that the estimator is asymptotically normal and suggested a valid full-sample bootstrap confidence interval to quantify the uncertainty. For the moderately extreme case, the paper showed that the limiting distribution of the estimator is no longer Gaussian, and proposed a $b$-out-of-$n$ bootstrap for valid inference. Note that, our proposed method is different from the existing methods, and it involves estimation of the conditional distributions (see Section \ref{Prop_method}). Recently, \citet{Litvinova_2020} established the asymptotic validity of the full-sample bootstrap for constructing confidence intervals for high-quantiles, tail probabilities, and other tail parameters of a univariate distribution. We employ the full sample bootstap considering the following steps to compute the standard error and confidence interval of the proposed estimator.
\begin{itemize}
	\item[Step 1.] Sample $(Y_{i}^{\ast},D_{i}^{\ast}, \boldsymbol{X_i^{\ast}})$ from the observed data $(Y_i,D_i, \boldsymbol{X_i})$, $i=1, \ldots,n$.
	\item[Step 2.] Based on the data $(Y_{i}^{\ast}, D_{i}^{\ast}, \boldsymbol{X_i^{\ast}})$, $i=1,\ldots,n$, obtain the estimates $\hat{\eta}_{\tau_0}^{\ast}$ and $\hat{\zeta}_{\tau_0}^{\ast}$.
	\item[Step 3.] Replicate Step 1 to Step 2 for $N_B$ number of times, where $N_B$ is a large number, say 1000.
	\item[Step 4.] Estimate the bias, standard error and confidence interval of $\hat{\eta}_{\tau_0}$ and $\hat{\zeta}_{\tau_0}$ from $N_B$ bootstrap estimates of $\hat{\eta}_{\tau_0}^{\ast}$ and $\hat{\zeta}_{\tau_0}^{\ast}$, respectively.
\end{itemize}

\section{Simulation study}\label{Sim} 
In this section, we generate synthetic data resembling the main features of the London traffic flow data set to compare different competing methods. It has been observed that the performance of the IPW-based method are very poor when the estimated propensity score are very small \citep{Diaz_2017, DR}. We seek to avoid such situations for a valid comparison. For this purpose, the parameters are chosen such that at least $25\%$ of the observations belong to the treated group. We consider both symmetric and skewed distributions to generate observations for the covariates. 
One of the two confounding variables, denoted by $X_1$ follows the normal distribution with mean 15 and standard deviation 6, and the other confounder $X_2$ follows the exponential distribution with mean 2. We then include a non-confounding variable $X_3$ generated from the normal distribution with both mean and standard deviation 1, and specify the following relationships between the response variable $Y$ and a binary treatment $D$:
$$\logit[\mathbb{E}(D|X_1,X_2,X_3)]=-3+0.1X_1+0.1X_2+0.2X_3$$
$$Y=10 + 15 D + X_1 + 3 X_2  + 2 X_1 D +(1+4 X_2 + 3 D)\epsilon,$$
where error $\epsilon$ is generated from two different symmetric distributions with mean 0. In the first case, we consider $\epsilon$ following the normal distribution with mean 0 and standard deviation 10 and in the second case $\epsilon$ follows the $t$-distribution with degrees of freedom 1. These two choices of error distribution allows us to compare the performance of the causal estimators for both light-tailed and heavy-tailed response variable. We generate each of the above data sets for two sample sizes $n=500$ and $n=1000$.

The transition point representing the shift from the bulk to the extreme part ($\tau_u$ in \eqref{decomposition}) is selected by the automated threshold selection method described in Section~\ref{sec:transition}. As a set of competing thresholds, we consider 10 quantiles with equally spaced probability levels ranging from 0.75 to 0.99. We then use quantile regression at each probability level $\tau$ with covariates as treatment $D$, and three more predictors, $X_1$, $X_2$ and $X_3$. The fitted quantiles are then used as covariate dependent thresholds for the GPD. The sequential multiple hypothesis approach based on Anderson-Darling goodness-of-fit test of \cite{bader2018} is then used to select the best threshold with given level $\lambda=0.05$. The selected threshold is used as the transition point and the corresponding probability level is taken as the transition point $\tau_u$. For evaluating the quantiles involved in the bulk part, as discussed in Section \ref{bulk}, we chose probability levels with an interval of 0.01 (mesh width). This choice of mesh width guarantees the required convergence rate to 0. We choose a total of 100 grid points in the computation of \eqref{sample-decomposition} with 75 and 25 equidistant points for computing the first (bulk part) and the second component (extreme part), respectively. For computing the bulk part of \eqref{sample-decomposition}, we adopt the  simultaneous quantile regression approach of \cite{Bondell_2010} and use the associated R package. In the extreme part of \eqref{sample-decomposition}, the GPD threshold is modelled with all the aforementioned covariates and the shape and scale parameters are estimated using the maximum likelihood method.

\begin{figure}[htp]
	\centering
	\subfigure  {
		\includegraphics[scale=0.35]{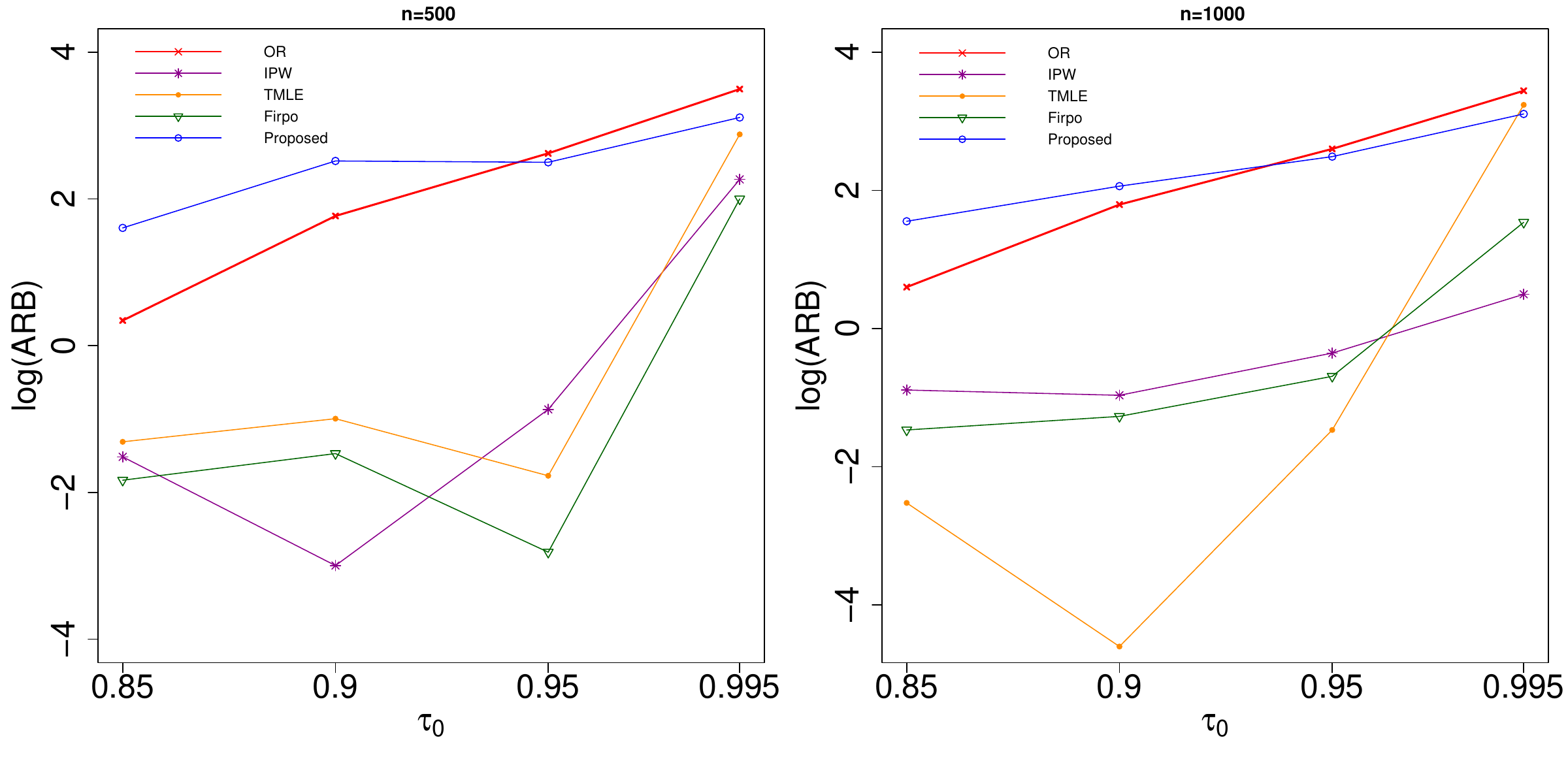}
		%		\label{fig:subfig01}
	}
	\subfigure {
		\includegraphics[scale=0.35]{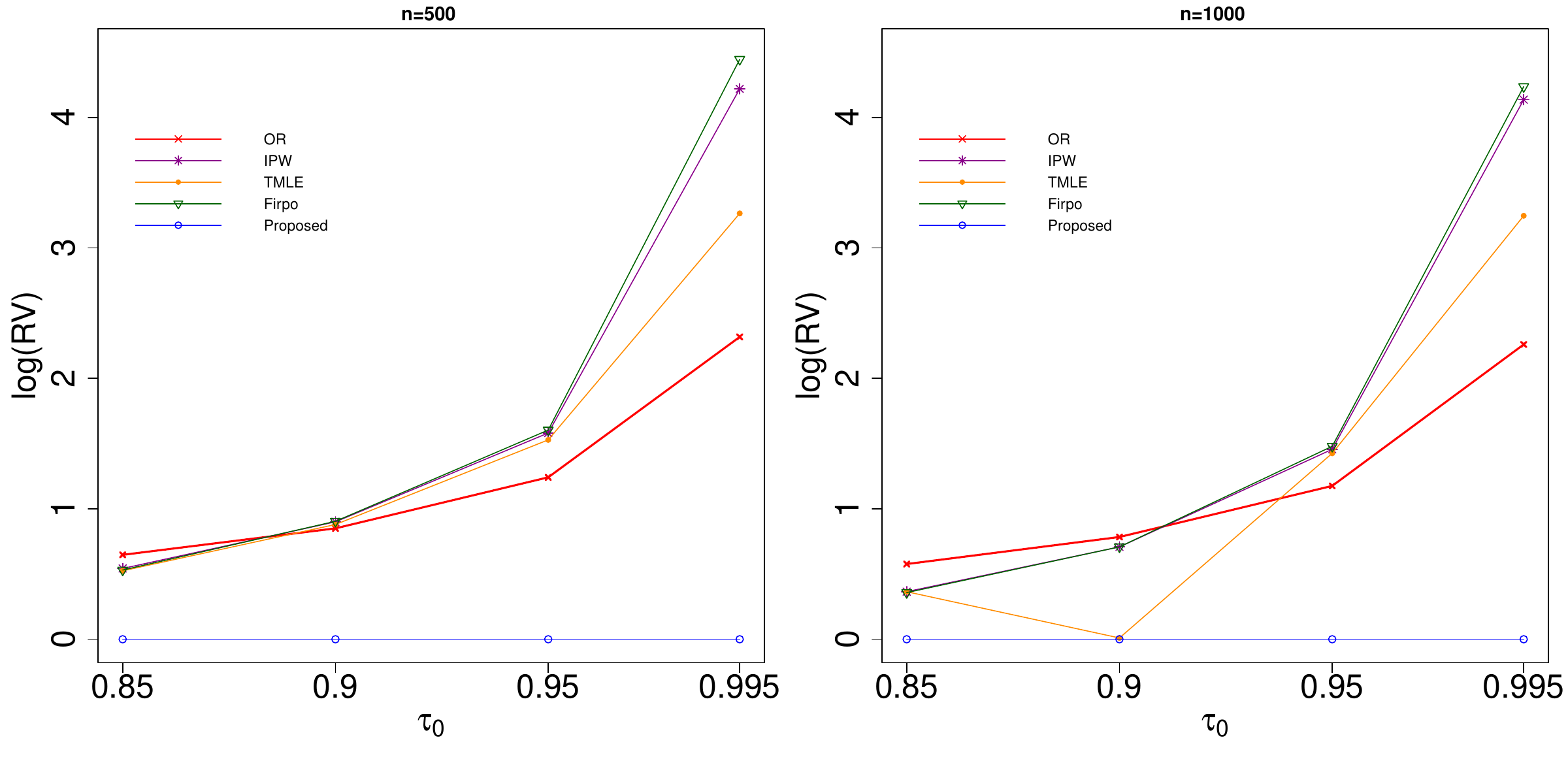}
		%		\label{fig:subfig02}
	}
	\subfigure {
		\includegraphics[scale=0.35]{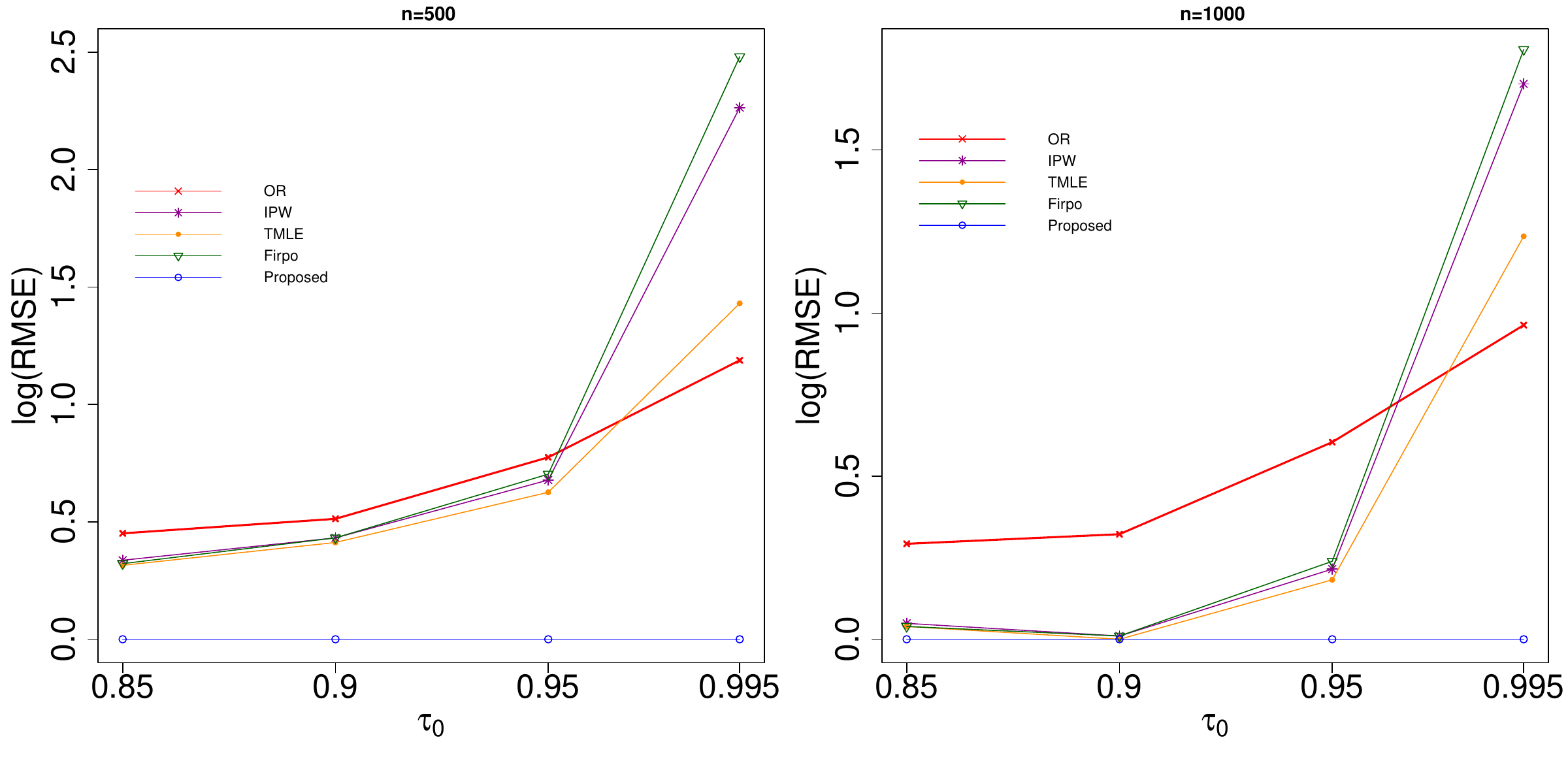}
		%		\label{fig:subfig03}		
	}
	\caption{Comparison of proposed estimator, OR, IPW, TMLE, and Firpo estimators for quantile treatment effect for the population at different probability levels based on the data generating process with Gaussian error. The top, middle and bottom rows show the comparison through three statistics: ARB, RV and RMSE, respectively, for two different sample sizes 500 and 1000. The numbers on the horizontal axis of each plot indicate the quantile levels, and five different symbols on the plot corresponds to five different estimators.}
	\label{tab1}
\end{figure}

\begin{figure}[htp]
	\centering
	\subfigure  {
		\includegraphics[scale=0.35]{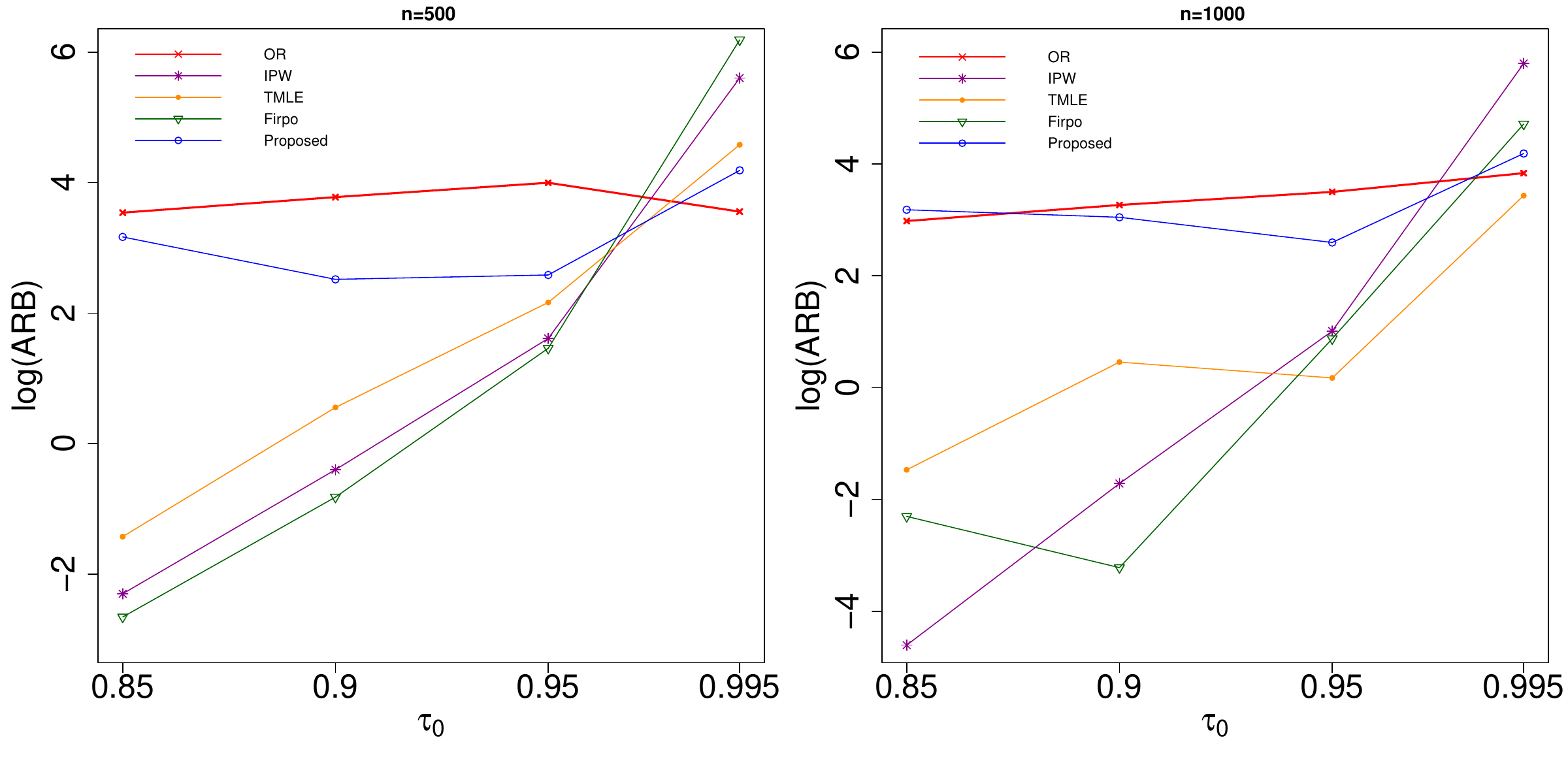}
		%		\label{fig:subfig01}
	}
	\subfigure {
		\includegraphics[scale=0.35]{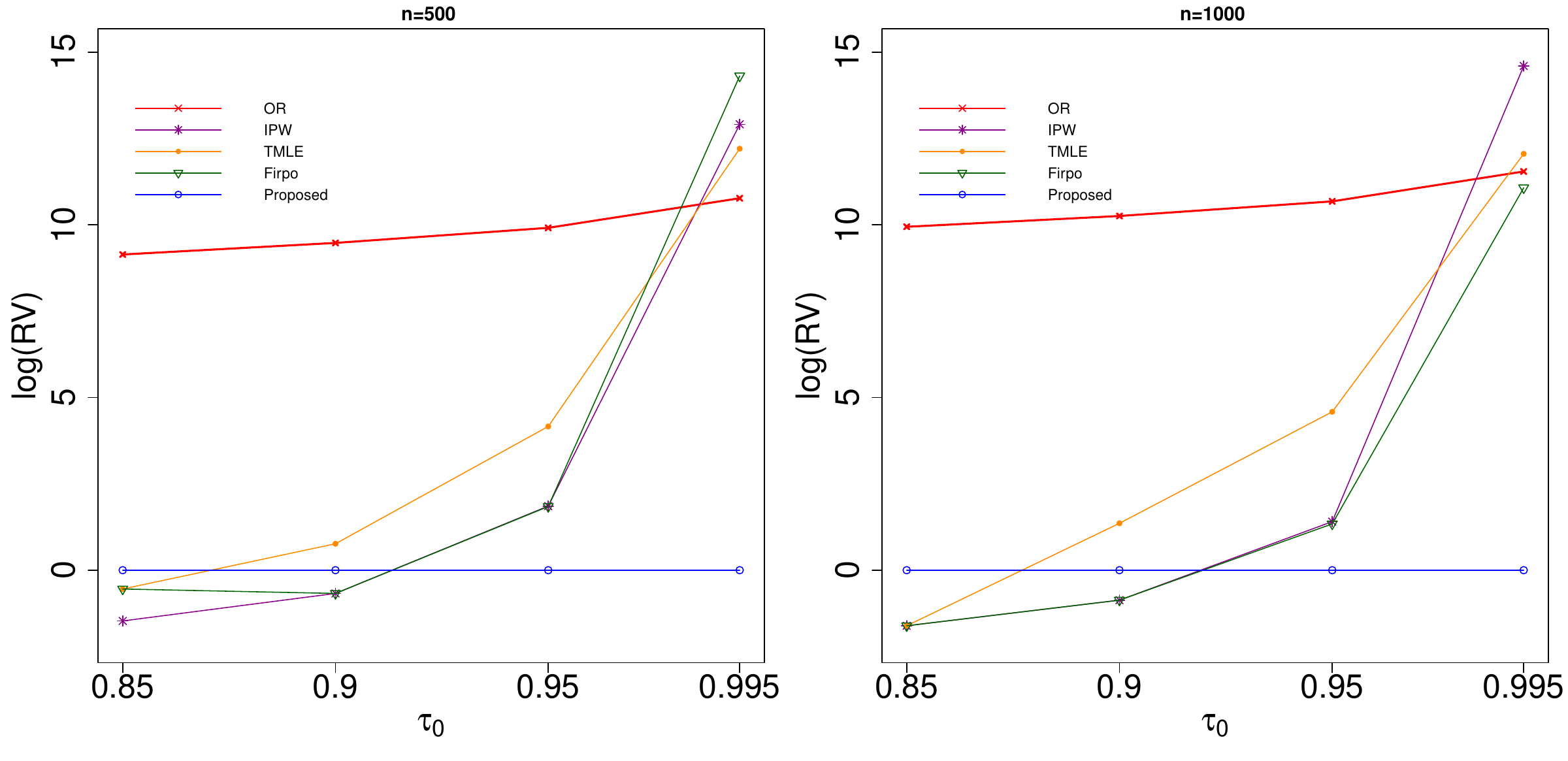}
		%		\label{fig:subfig02}
	}
	\subfigure {
		\includegraphics[scale=0.35]{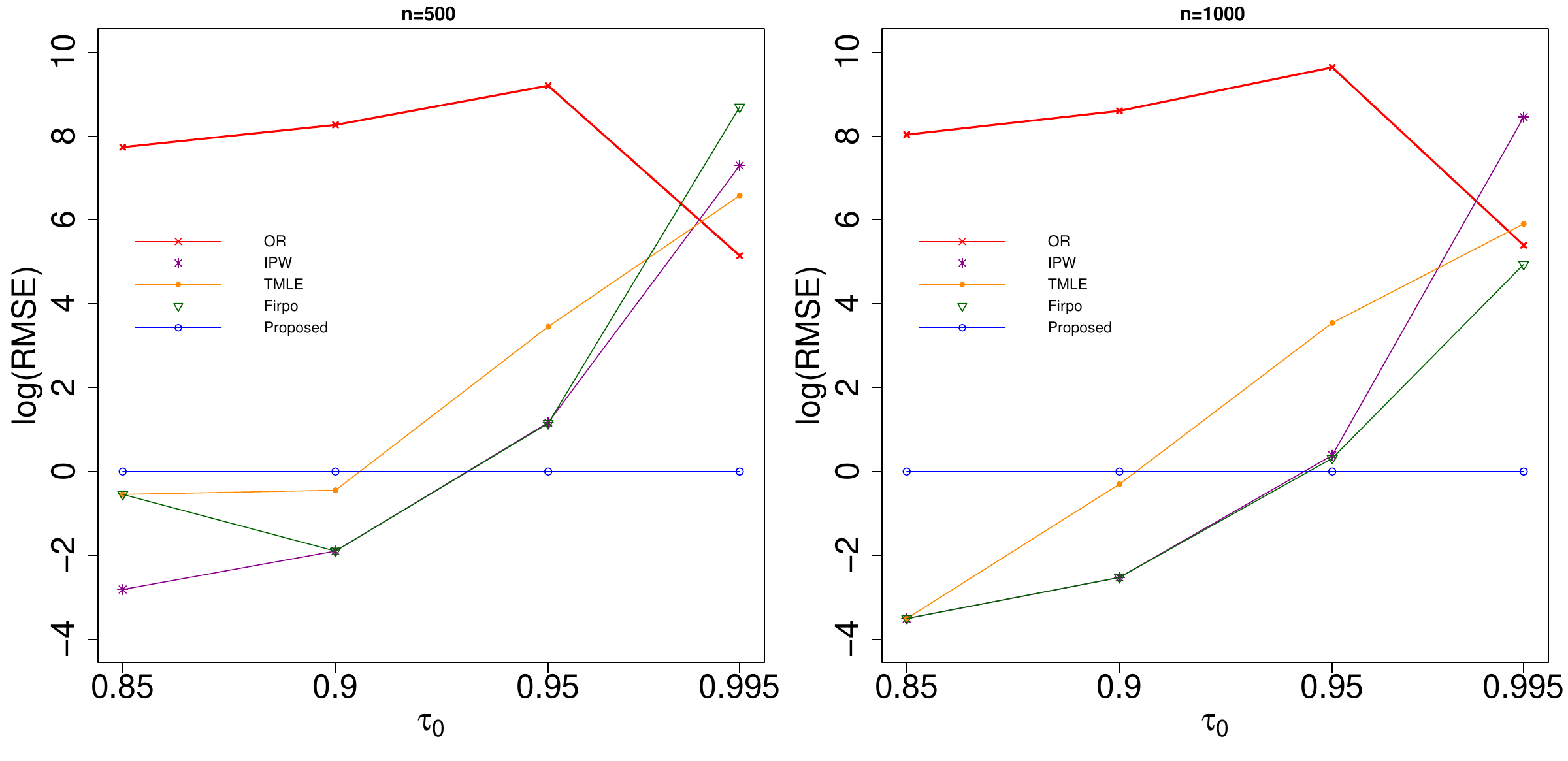}
		%		\label{fig:subfig03}		
	}
	\caption{Comparison of proposed estimator, OR, IPW, TMLE, and Firpo estimators for quantile treatment effect for the population at different probability levels based on the data generating process with heavy-tailed error from t-distribution. The top, middle and bottom rows show the comparison through three statistics: ARB, RV and RMSE, respectively, for two different sample sizes 500 and 1000. The numbers on the horizontal axis of each plot indicate the quantile levels, and five different symbols on the plot corresponds to five different estimators.}
	\label{tab2}
\end{figure}

\begin{figure}[htp]
	\centering
	\subfigure  {
		\includegraphics[scale=0.35]{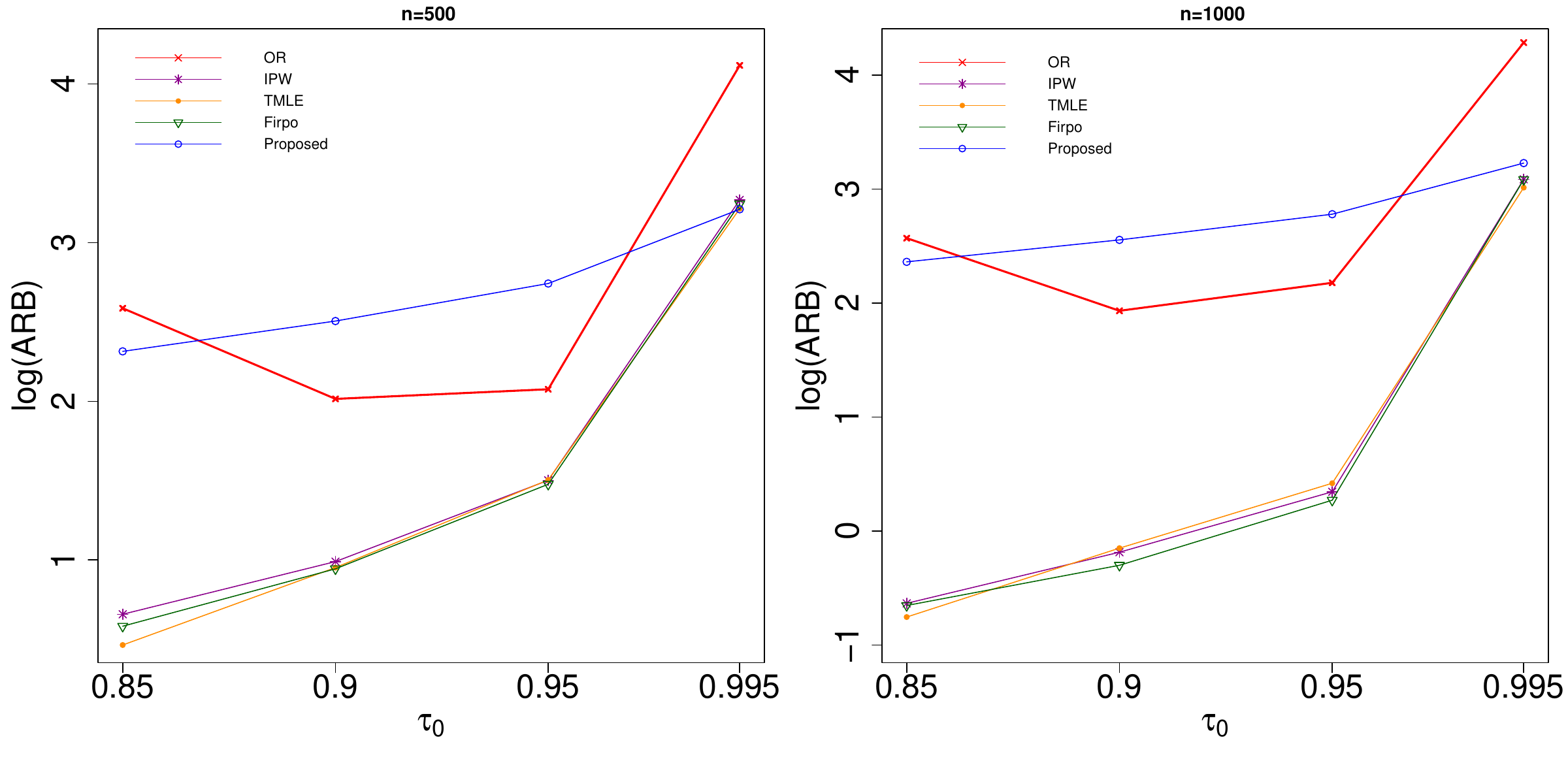}
		%		\label{fig:subfig01}
	}
	\subfigure {
		\includegraphics[scale=0.35]{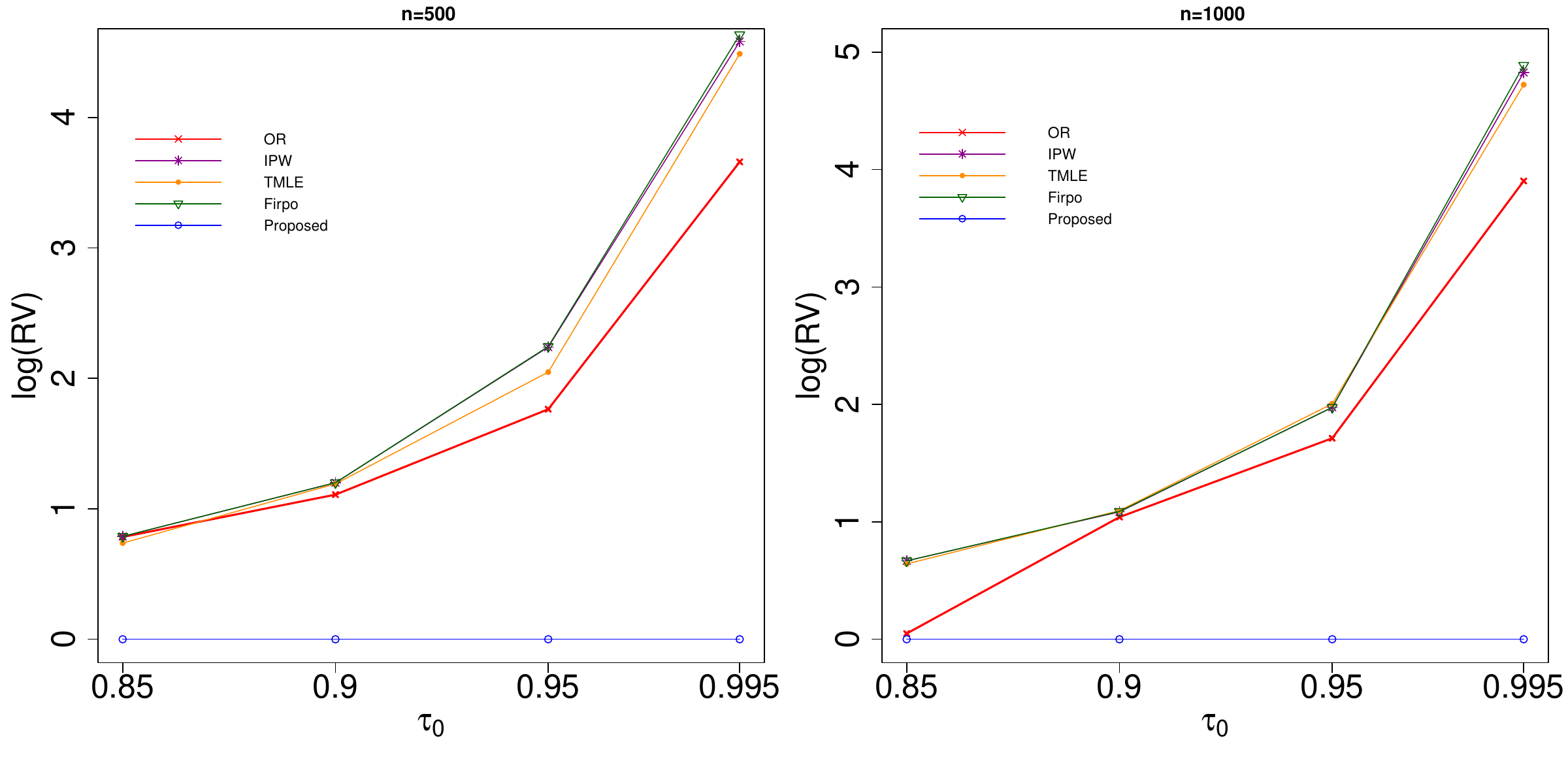}
		%		\label{fig:subfig02}
	}
	\subfigure {
		\includegraphics[scale=0.35]{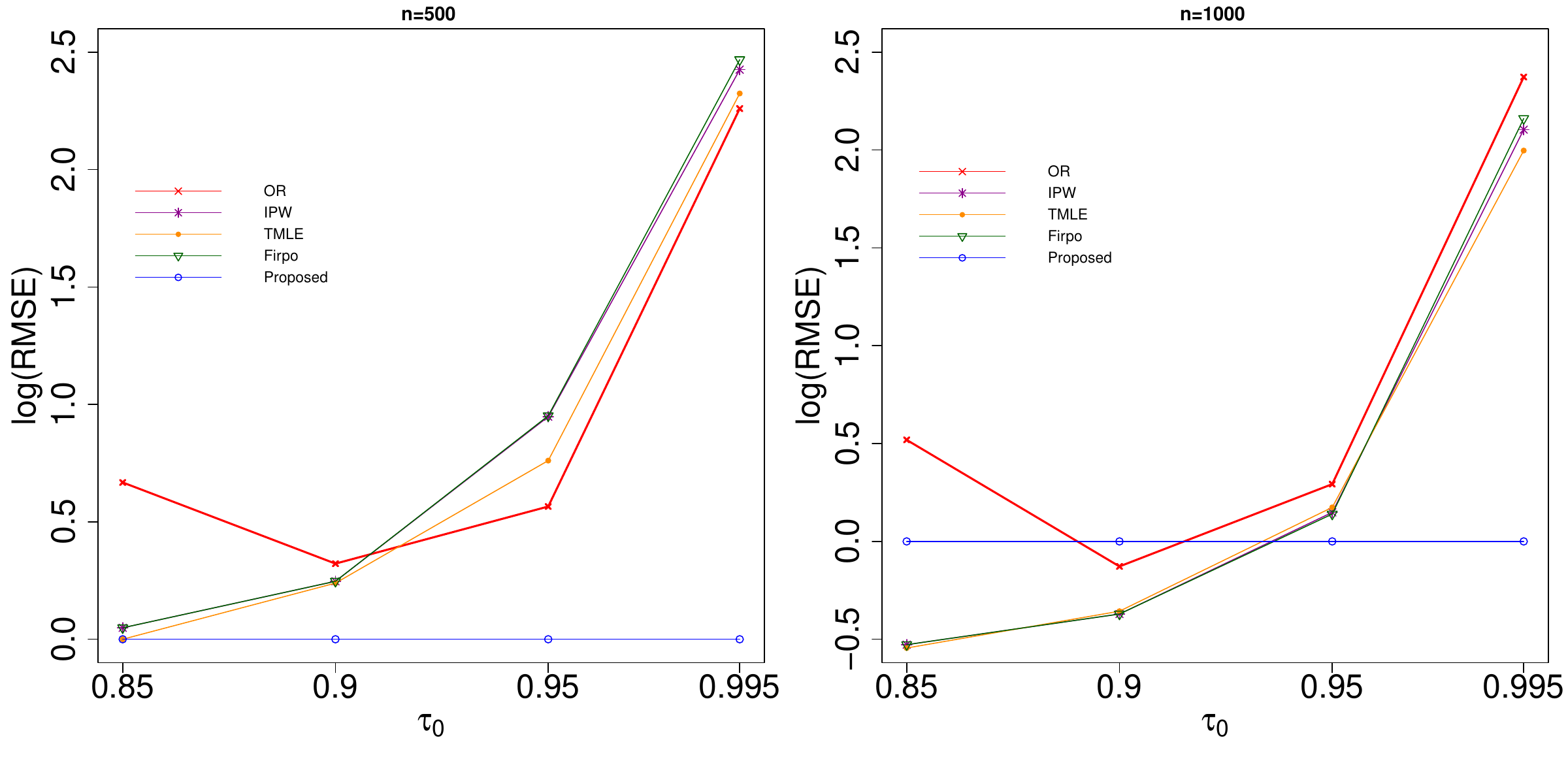}
		%		\label{fig:subfig03}		
	}
	\caption{Comparison of proposed semiparametric, OR, IPW, TMLE, and Firpo estimators for quantile treatment effect for the treated at different probability levels based on the data generating process with Gaussian error. The top, middle and bottom rows show the comparison through three statistics: ARB, RV and RMSE,  respectively, for two different sample sizes 500 and 1000. The numbers on the horizontal axis of each plot indicate the quantile levels, and five different symbols on the plot corresponds to five different estimators.}
	\label{tab3}
\end{figure}

\begin{figure}[htp]
	\centering
	\subfigure  {
		\includegraphics[scale=0.35]{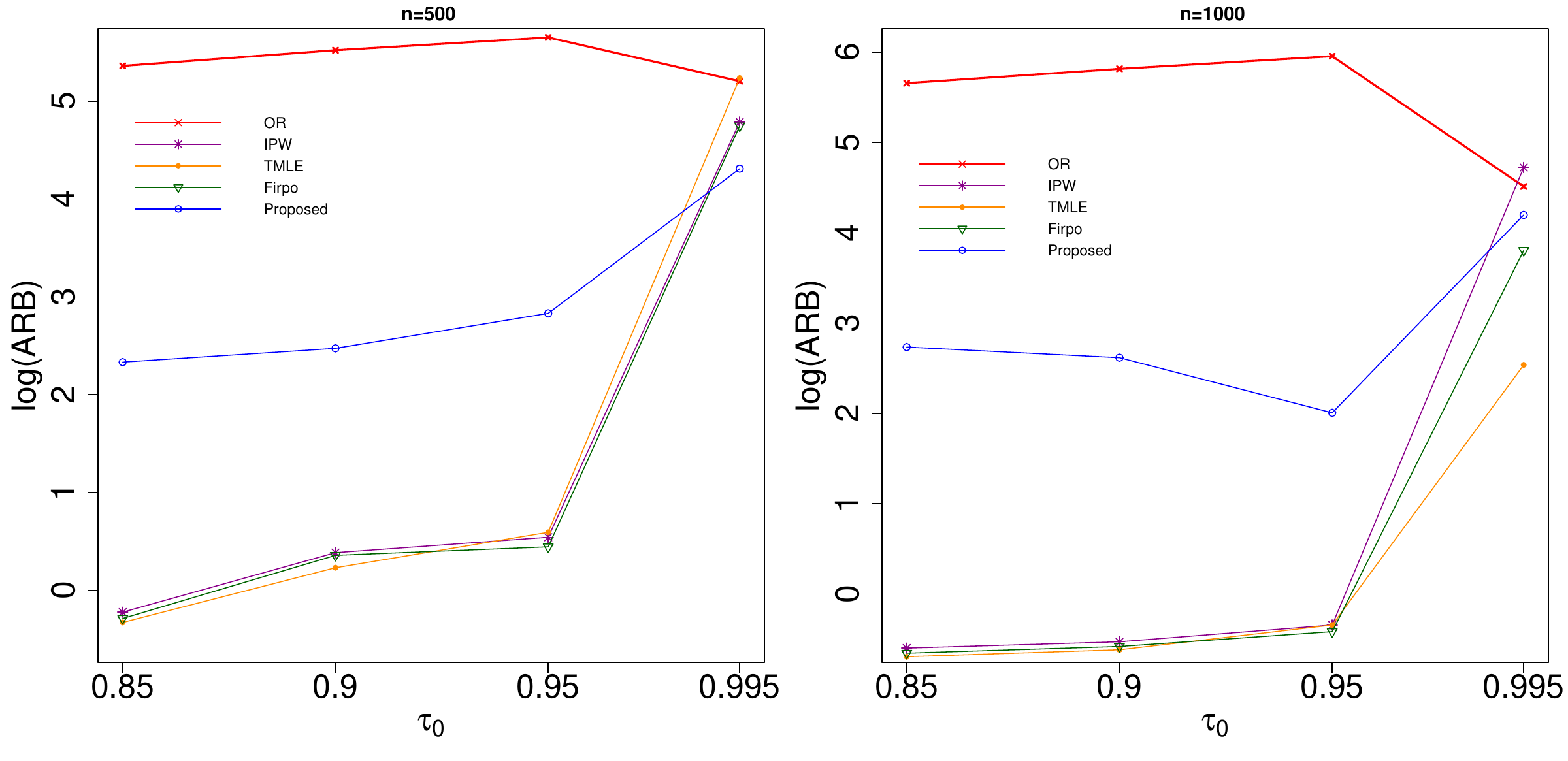}
		%		\label{fig:subfig01}
	}
	\subfigure {
		\includegraphics[scale=0.35]{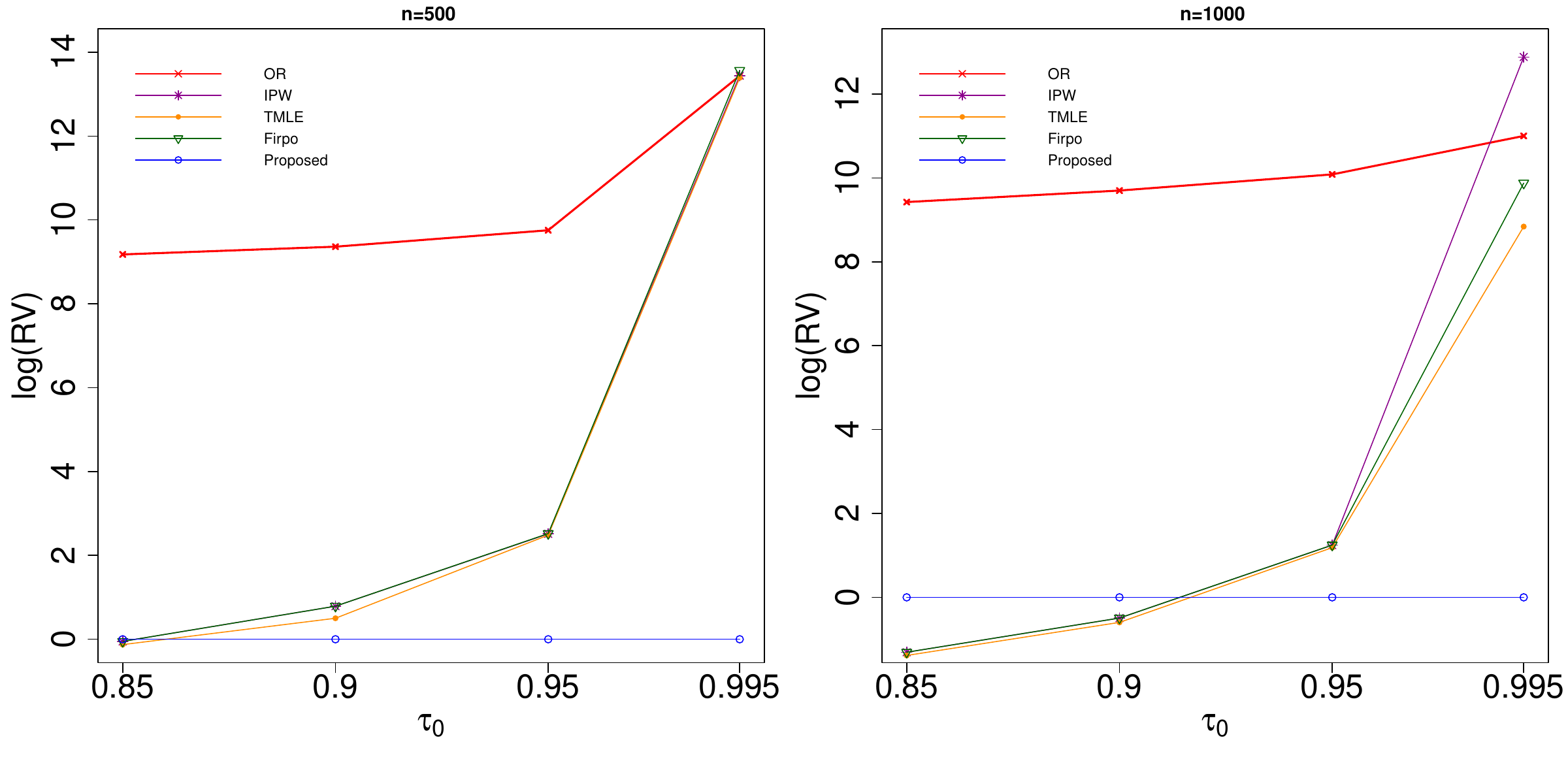}
		%		\label{fig:subfig02}
	}
	\subfigure {
		\includegraphics[scale=0.35]{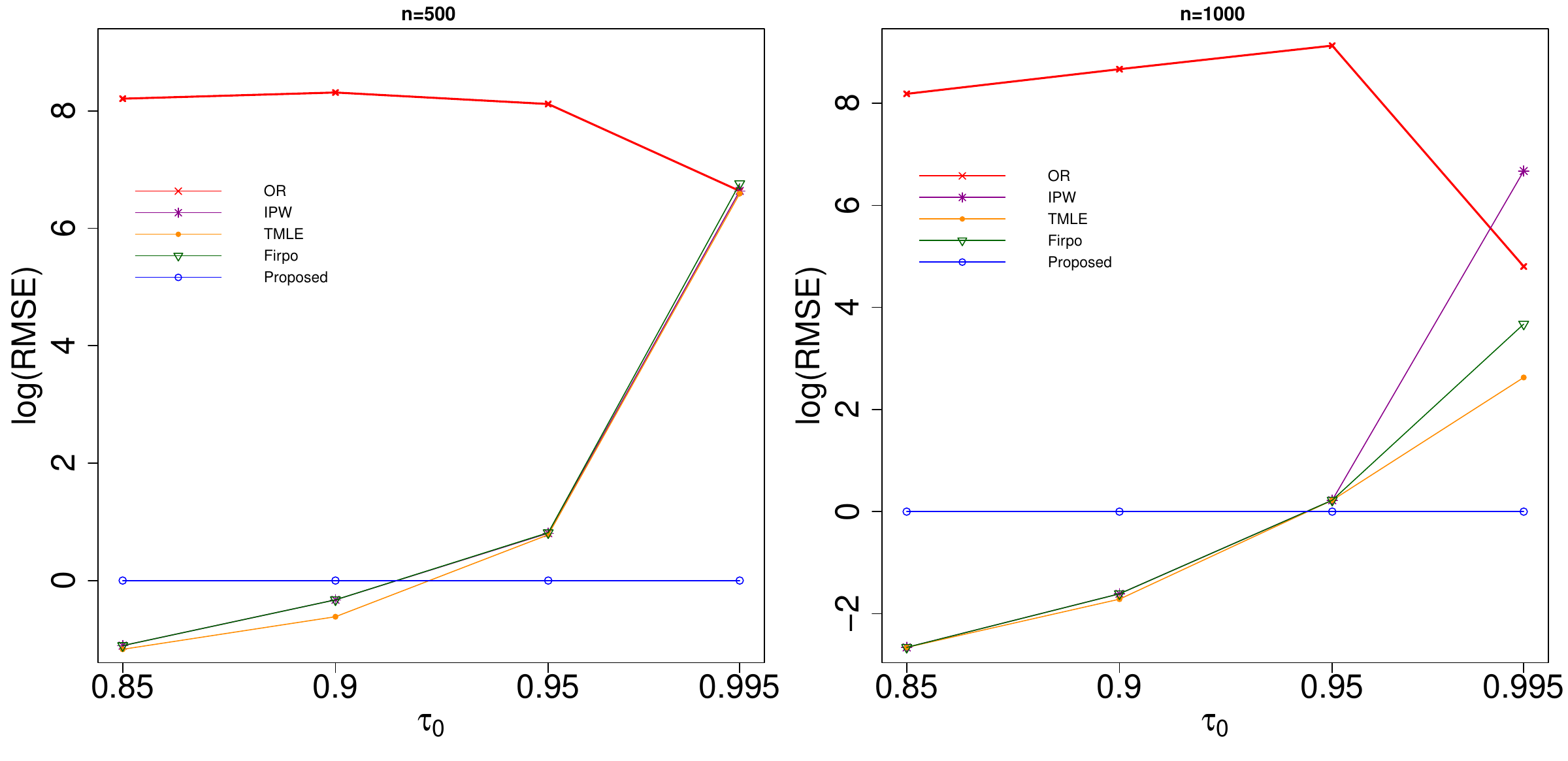}
		%		\label{fig:subfig03}		
	}
	\caption{Comparison of proposed semiparametric, OR, IPW, TMLE, and Firpo estimators for quantile treatment effect for the treated at different probability levels based on the data generating process with heavy-tailed error from t-distribution. The top, middle and bottom rows show the comparison through three statistics: ARB, RV and RMSE, respectively, for two different sample sizes 500 and 1000. The numbers on the horizontal axis of each plot indicate the quantile levels, and five different symbols on the plot corresponds to five different estimators.}
	\label{tab4}
\end{figure}

We compare the performance of the proposed estimator with the existing estimators for both population and treated subpopulation.
As a set of competing estimators, we consider the targeted maximum likelihood estimate (TMLE) of \citet{Diaz_2017} and the estimator of \citet{Firpo_2017} along with the OR and IPW estimates by \citet{zhang2012}. All these methods are based on correctly specified OR and/or PS models. The competing estimators are then compared based on 1000 iterations with respect to the following statistics: the absolute relative bias (ARB), the relative variance (RV) and relative mean squared error (RMSE).
More precisely, we compute the following statistics with $L$ iterations: 
$$ ARB = \Bigg\lvert\frac{1}{L}\sum_{i=1}^L \frac{\hat \eta^{(i)}_{\tau_0}-\eta_{\tau_0}}{\eta_{\tau_0}}\Bigg\rvert,\; \;
RV=\frac{\sum_{i=1}^L \left[\hat{\hat \eta}^{(i)}_{\tau_0}-\frac{1}{L}\sum_{i=1}^L\hat{\hat \eta}^{(i)}_{\tau_0}\right]^{2}}{\sum_{i=1}^L \left[\hat \eta^{(i)}_{\tau_0}-\frac{1}{L}\sum_{i=1}^L\hat \eta^{(i)}_{\tau_0}\right]^{2}},\;\;
RMSE=\frac{\sum_{i=1}^L \left[\hat{\hat \eta}^{(i)}_{\tau_0}-\eta_{\tau_0}\right]^{2}}{\sum_{i=1}^L \left[\hat \eta^{(i)}_{\tau_0}-\eta_{\tau_0}\right]^{2}},
$$ where $\hat \eta^{(i)}_{\tau_0}$ and $\hat{\hat \eta}^{(i)}_{\tau_0}$ are the estimates of the population quantile effect at $\tau_0$ based on the proposed method and a competitor in the $i$th iteration, respectively, with $L=1000$. Similarly, we compute these metrics for the treated subpopulation (i.e., for $\zeta_{\tau_0}$).
The results for the quantile effects on population are summarized graphically in Figures~\ref{tab1}-\ref{tab2} based on the aforementioned statistics for all the estimators under consideration. The values for all the statistics are presented on a natural log-scale. Similar comparison are performed for the quantile treatment effects on the treated subpopulation and the findings are presented in the  Figures~\ref{tab3}-\ref{tab4}.

First, we discuss the results corresponding to quantile causal effects for the population (Figures~\ref{tab1}-\ref{tab2}). The ARB associated with IPW, TMLE and Firpo estimates are small for all the probability levels except at $\tau_{0}=0.995$. The ARB increases significantly for all the estimates at probability level $\tau_{0}=0.995$. In particular, the ARB of IPW and Firpo estimates are turns out to be very large compared to the other methods at probability level $\tau_{0}=0.995$ when the data generation mechanism involves a heavy-tailed error distribution (see Figure~\ref{tab2}). For all the probability levels, the 
RV and RMSE of the proposed method are smaller compared to all other methods when the data generation mechanism involves a normal error distribution (see Figure~\ref{tab1}). Similar patterns are also observed at $\tau_{0}=0.95$ when the data generation mechanism involves a heavy-tailed error distribution (see Figure~\ref{tab2}). In this setup, the RV and RMSE of all the competing estimates are exorbitantly high compared to the proposed method at  $\tau_{0}=0.995$. Next we consider the results corresponding to quantile causal effects for the treated (see Figures ~\ref{tab3}-\ref{tab4}). Here, the findings are very similar to the case of causal effects on the population except for a few notable differences. The performance of all estimates except the OR method at $\tau_{0}=0.995$ are comparable with respect to the ARB when the data generation mechanism involves a normal error distribution (see Figure ~\ref{tab3}). In contrast, when the data generation mechanism involves a heavy-tailed error distribution, the ARB of the proposed method is considerably smaller compared to all other competitive methods for $n=500$. The RV and RMSE of all the competitive estimates are considerably higher compared to the proposed method at probability level $\tau_{0}=0.95$ and $\tau_{0}=0.995$. This pattern is more dominant when the data generation mechanism involves a heavy-tailed error distribution (see Figure~\ref{tab4}). These findings for the above simulation study demonstrate the effectiveness of the proposed method over the existing methods for estimating the causal effect at extreme quantiles.

\section{Analysis of London cycle superhighways effects}\label{Application}
\begin{figure}[htp]	
	\centering
	\includegraphics[scale=0.5]{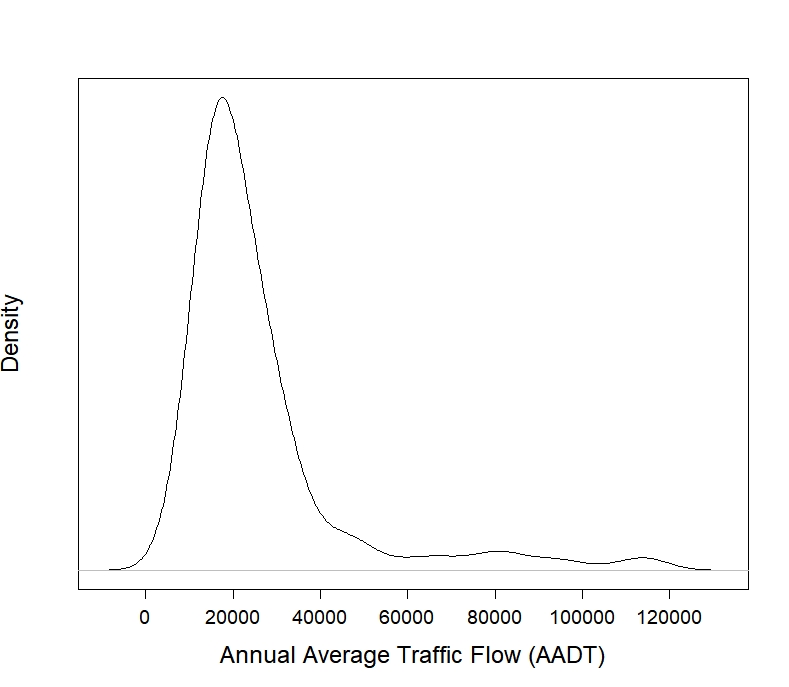}
	\caption{Density plot of Annual Average Traffic Flow during 2011-2014 in London metropolitan area.}
	\label{density-aadt}
\end{figure}

The primary focus of this study is to evaluate the effect of London cycle superhighways (intervention) to contain extreme congestion. %If the overall traffic volume goes up but the road space is being shared with other forms of transport, then there is less space for the mobility of cars, which impacts the traffic speed on the road and therefore congestion increases \citep{Yuan2015}. 
The data that are available and the 
rationale to construct the response (Annual Average Daily Traffic) and covariates are described below.
\begin{itemize}
	\item[(a)] Annual Average Daily Traffic (AADT): the total volume of vehicle traffic of a highway or a road in a year divided by 365 days. To measure AADT on individual road segments, traffic data is collected by an automated traffic counter, hiring an observer to record traffic or licensing estimated counts from GPS data providers. AADT is a simple but useful measurement to indicate busyness of a road.
	%	\item[(b)] Traffic speed -- calculated using time-mean-speed method based on the individual speed records for vehicles passing a point over a selected time period. Speed is also a fundamental measurement in transport engineering and used for maintaining a designated level of service.
	\item[(b)] Total Cycle Collisions (TCS): total number of cycle collisions causing injury based on police records from the STATS19 accident reporting form and collected by the UK Department for Transport. The location of an accident is recorded using coordinates which are in accordance with the British National Grid coordinate system. The CS routes were intended to reduce the risk of accidents for cyclists and the route allocation is possibly influenced by TCS. It is expected that the accident rates will affect the traffic characteristics.
	\item[(c)] Bus-stop density: the ratio of the number of bus-stops to the road length. The presence of bus-stops is expected to affect the traffic flow and speed due to frequent bus-stops and pedestrian activities. The allocation of CS routes were designed to avoid areas with high bus-stop density for safety of the cyclists.
	\item[(d)] Road network density: with the available geographical information system we could also represent the road network density in each zone by using a measure of the number of network nodes per unit of area. A network node is defined as the meeting point of two or more links. To safeguard from conflicting turning movements the CS paths are routed through the areas with high road network density.  
	%To provide safe passage for cyclists, CS paths are designed to avoid conflicting turning movements which is common in the areas with high road network density.   
	\item[(e)] Road length: high capacity networks tend to depress land values which in turn will influence the socio-economic profile of the people who live close together. Data for road length for each zone was generated using geographical information system software.
	\item[(f)] Road type: a binary variable where `1' represents dual-carriageway and `0' represents single-carriage. This is an important feature since we might expect traffic congestion in single-carriage roads.
	\item[(g)] Density of domestic buildings: this is a potentially useful feature since we might expect congestion to be associated with the nature of land use and the degree of urbanization. Also, the allocation of the CS paths are possibly influenced by land use characteristics.
	\item[(h)] Density of non-domestic buildings: rising housing costs in business and office districts force people to live further away, lengthening commutes, and affecting traffic flow and speed. As mentioned before, this feature may influence allocation of the CS paths.
	\item[(i)] Road area density: the ratio of the area of the zone's total road network to the land area of the zone. The road network includes all roads in the including motorways, highways, main or national roads, secondary or regional roads. It is expected that the traffic flow is associated with road density. 
	\item[(j)] Employment density: traffic generation potential depends on economic activity and we proxy this by employment density. High employment density tends to influence pedestrian activity which in turn affects traffic speed. The CS paths are designed to provide coverage in the areas with high employment density and encourage commuters to use cycling as a regular mode of transport. 
\end{itemize}
Figure~\ref{density-aadt} shows density plot of AADT, exhibiting heavy-tailed nature of the response variable. 
This feature of the response variable and total sample size of 450 observations is adequate to apply extreme theory and related inference \citep{Hosking1987}. 
Using the proposed method (with same set of grid points and level of test as used in the preceding simulation study) we estimate the quantile treatment effects for the population and for the treated at different intermediate and extreme quantiles corresponding to probability levels $\tau_{0}$=0.85, 0.90, 0.95 and 0.995. We consider the OR, IPW, TMLE and Firpo methods along with the proposed method for the analysis. Propensity score is estimated using a logit model with the following factors: total cycle collision, bus-stop density, road  network  density, road length, density  of  domestic  buildings, density of non-domestic buildings, road area density, and employment density. The OR method by \citet{zhang2012} and the proposed method use density of domestic  buildings, density of non-domestic  buildings, road area density, road network density, road type, bus-stop density as covariates. The estimated quantile treatment effects of CS on annual average daily traffic (AADT) relative to the respective quantile of AADT in the pre-intervention period are presented in Tables \ref{R_QTE}-\ref{R_QTT}. The pre-intervention (2007 to 2010) quantiles of AADT are computed based on the observations from the same locations as considered for the post-intervention (2011-2014) period.
The standard errors (SEs) and the corresponding 95\% confidence intervals (CI) for all the estimates, with the exception of Firpo estimate at probability level $\tau_{0}=0.995$, are obtained from 1000 full bootstrap samples. To find the SE and CI of the causal effects based on Firpo method at probability level $\tau_{0}=0.995$, we employ $b$-out-of-$n$ bootstrap proposed by \citet{Zhang_2021}. See Section \ref{Boot} for details. 

The result based on the proposed estimate indicate roughly 29\% increase in traffic flow compared to the pre-intervention period at probability levels $\tau_{0}=0.995$ both for the entire city and the treated locations. The 95\% bootstrap CI for the proposed estimate exclude $0$, suggesting a significant effect of CS on AADT. However, as observed in the simulation study, the SE of the estimated causal effects of CS at probability level $\tau_{0}=0.995$ based on OR, IPW, and Firpo methods are very high and the most of the results are not significant except the estimate based on TMLE. The SE of the proposed method is significantly lower than those of the OR, IPW, TMLE and Firpo methods which is also reflected in the width of the confidence intervals at all probability levels for treated. A similar pattern is also observed for the population estimates at probability levels $\tau_{0}=0.95$ and $\tau_{0}=0.995$. 

Overall, our analysis indicates that the introduction of CS can trigger extreme traffic flows and increase traffic mobility. The protected cycleways in London including CS is just over 1\% of the capital’s roads and CS is only a quarter of the protected cycleways. In 2020, the five busiest roads in the U.K. were in London, however, none of these five roads is within proximity of cycleways. 
%It is important to note that such observations only reflect an average scenario rather than the extreme cases.  
The U.K. has 38.3 million registered motor vehicles, up from 27 million in 2007. With cities such as London not appreciably adding road capacity with a massive expansion of motor vehicles will lead to traffic congestion \citep{Forbes}. It is not unexpected that several unobserved factors associated with the transport network and other interventions play a crucial role in mitigating potential traffic problems anticipated by the introduction of cycle lanes. The results from our analysis provide more insights, but for a comprehensive understanding to formulate effective transport strategies, the effect of the cycleways on traffic speed requires further investigation. 
%It is important that {\bf state-of-the-art methods combining tools from different domains} are applied in such important applications and given the superiority of our methods over existing competitors provide a framework for analysis of high consequence events.

\begin{table}
	\caption{\label{R_QTE} The quantile causal effect of CS on AADT relative to the respective quantile of AADT in the pre-intervention period for population  at different probability levels $\tau_{0}$.}	
	\centering
	%\resizebox{\textwidth}{50mm}{
	\begin{centering}
		\begin{tabular}
			{cccHHcc}\\\hline
			$\tau_{0}$ & Method &  Estimate (\%) & Corrected Estimate (\%)  & Mean  & SE & 95\% CI\\ \hline\hline
			& OR & 43.34 & 44.58 & 59.97 &  11.50 &  (25.36, 70.85) \\
			&  IPW & 51.41 & 22.47  & 80.36 & 73.06 &  (-3.27,  219.09) \\
			0.85	  & TMLE & 38.08 & - & - &  26.85 & (4.47, 135.68) 
			\\
			& Firpo & 108.17 & 84.27 & 132.09 &  22.37 & (41.85, 135.01) 
			\\
			& Proposed & 98.52 & 93.95 & 103.08 &  19.76 & (65.19, 135.09) 
			\\\hline
			& OR & 41.33 & 40.13 & 42.52 &  11.18 &  (22.88, 65.65) \\
			&  IPW & 128.95 & 160.35 & 97.55 & 51.70 &  (0.35,101.92) \\
			0.90	 & TMLE & 100.47 & - & - &  38.50 & (3.24, 98.34) 
			\\
			& Firpo & 121.26 & 148.22 & 94.30 & 15.39 & (30.40, 97.85) 
			\\
			& Proposed & 71.60 & 74.77 & 68.42 & 14.40 & (47.07, 97.63) 
			\\\hline
			& OR & 37.60 & 36.84 & 38.36 &  10.43 &  (19.07, 59.49) \\
			&  IPW & 61.50 & 65.36 & 57.63 & 19.66 &  (0.31,  83.81) \\
			0.95	  & TMLE & 55.03 & - & - &  12.15 & (2.03, 61.76) \\
			& Firpo & 59.69 & 62.11  & 59.78 & 9.71 & (19.21, 61.45) 
			\\
			& Proposed & 44.98 & 43.52  & 46.44 & 8.91 & (28.57, 60.86) 
			\\\hline
			& OR & 36.71 & 42.87 & 30.53 &  20.81 &  (-9.16, 69.66) \\
			&  IPW & 2.08 & -2.49 & 6.65 & 9.32 &  (-4.78, 22.92) \\
			0.995	 & TMLE & 15.37 & - & - &  6.26 & (1.32, 39.97) \\
			& Firpo & -0.95 & -40.42  & 38.52 & 15.88 & (-19.21, 40.90) 
			\\
			& Proposed & 29.11 & 27.93 & 30.29 & 5.86 & (19.31, 39.77)
			\\\hline
		\end{tabular}
	\end{centering}
	%	}
\end{table}

\begin{table}
	\caption{\label{R_QTT} The quantile causal effect of CS on AADT relative to the respective quantile of AADT in the pre-intervention period for treated at different probability levels $\tau_0$.}
	\centering
	%\resizebox{\textwidth}{50mm}{
	\begin{centering}
		\begin{tabular}
			{cccHHcc}\\\hline
			$\tau_{0}$ & Method &  Estimate (\%) &  Corrected Estimate (\%)  & Mean  & SE & 95\% CI\\ \hline\hline
			& OR & 41.27 & 39.39 & 43.14 & 15.90 &  (14.56, 71.44) \\
			&  IPW & 49.69 & 49.27 & 50.11 & 17.10 &  (21.55,  81.00) \\
			0.85	  & TMLE & 49.22 & - & - &  20.75 & (2.56, 77.02) \\
			& Firpo & 50.80 & 22.78 & 78.82 & 23.11 & (26.97, 99.01) 
			\\
			& Proposed & 44.65 & 41.96 & 47.33 & 8.76 & (30.56, 62.12) 
			\\\hline
			& OR & 50.27 & 54.64 & 45.90 & 10.09 &  (21.61, 62.42) \\
			&  IPW & 60.94 & 70.52 & 51.36 & 13.51 &  (19.69, 71.28) \\
			0.90	  & TMLE & 55.43 & - & - &  18.22 & (-2.32, 66.11) \\
			& Firpo & 67.85 & 70.51 & 65.18 & 19.45 & (18.78, 82.99) 
			\\
			& Proposed & 37.71 & 35.53 & 39.88 & 7.27 & (25.95, 52.19) 
			\\\hline
			& OR & 45.04 & 43.38 & 46.69 &  8.65 &  (26.46, 59.49) \\
			&  IPW & 38.19 & 35.06 & 41.31 & 13.46 &  (14.21,  69.27) \\
			0.95	  & TMLE & 31.07 & - & - &  14.33 & (-3.63, 56.57) \\
			& Firpo & 71.63 & 82.54 & 60.71 & 17.82 & (20.72, 76.37) 
			\\
			& Proposed & 34.31  & 36.42 & 32.21 &  7.04 & (23.06, 48.46) 
			\\\hline
			& OR &  -8.51 & -14.54 & -2.48 & 25.50 &  (-80.79,  21.02) \\
			&  IPW & 15.53 & 19.00 & 12.06 & 8.79 &  (-2.37, 26.51) \\
			0.995	  & TMLE & -3.60 & - & - &  12.32 & (-22.56, 24.75) \\
			& Firpo & 17.48 & -14.24 & 49.20 & 15.62 & (3.08, 63.71)
			\\
			& Proposed & 28.80 & 30.40 & 27.18 & 5.90 & (19.27, 40.13)
			\\\hline
		\end{tabular}
	\end{centering}	
	%		 }
\end{table}

\section{Discussions}\label{Discussion}
In this article we have introduced a methodology that can be used to draw inference for causal effects at extreme quantiles. We have modelled the relationships between the covariates and the response and use those relationships to predict the outcome for both the treatment statuses and to estimate the treatment effect at certain extreme quantiles. The key methodological contribution is the improvisation of the conventional outcome regression (OR) model to combine a semi-parametric quantile regression framework with a heavy-tailed parametric component for the extreme tail of the response distribution. This approach also addresses issues with model misspecification due to parametric assumptions and accounts for the high variability at the extreme tails of the distribution.
The inverse propensity weighted (IPW) estimate of \cite{zhang2012} and the estimator proposed by \citet{Firpo_2017} are the usual alternatives for estimating causal effects at intermediate quantiles of a heavy-tailed distribution that avoid some of the parametric assumptions associated with the regression function and/or link function embedded in the OR approach. But these estimators exhibit high volatility at the extreme tails similar to that of the OR method, whereas the proposed method performs  substantially better than all the competitors under consideration. 
Recently, \citet{Deuber2022} considered the estimator proposed by \citet{Firpo_2017} and adjusted it with the Hill estimator for the extreme value indices to estimate the extremal quantile treatment effect. The Hill estimator depends on the choice of selection of the number of upper order statistics $\kappa$ such that $\kappa\rightarrow \infty$ and $\kappa/n\rightarrow 0$. The choice of $k$ is a crucial issue and several practical choices are prescribed in the literature. Our proposed method is different from this method and it doesn't require for selection the upper order statistics. Also, the performance of the Hill estimator based method is very similar to the method proposed by \citet{Firpo_2017} with respect to mean squared error.
%It would be an interesting problem to utilize the Hill estimator in our framework and develop a new methodology for estimating the causal extremal quantile treatment effects.

In observational studies, one can never be sure that a model for the treatment assignment mechanism or an outcome regression model are correct. An alternative approach is to develop a doubly-robust (DR) estimator. Several DR estimation methodologies are proposed in the literature \citep{DR}. The approach suggested by \citet{zhang2012} for doubly robust estimation for quantile causal effects is not readily extendable to our case. However, one can use augmented regression methods considering a suitable function of the propensity score $\phi(\pi(D_i|x_i, \hat{\gamma}))$ as an additional covariate in (\ref{constrained1}) and (\ref{gpd-est}). Although empirical validation of the DR property seems possible, the study of theoretical properties is a challenging problem under the current set-up. One can further develop a new doubly robust estimation methodology based on the targeted maximum likelihood approach considered by \citet{Diaz_2017}. 

We have used the proposed method to analyse the effect of London Cycle Superhighways. Our results suggest that the introduction of Cycle Superhighways can increase extreme traffic flow, encouraging further research and cautious approaches for any possible extension of Cycle Superhighways in metropolitan cities like London, which are heavily affected by congestion. In this context, it would be an interesting problem to develop a statistical framework that can be used to derive inference for quantile causal effects incorporating both pre-intervention and post-intervention data as discussed in \citet{Bhuyan2021}. Also, it is of great interest for general public and epidemiologists, particularly in light of the increase in cycling and the recent proliferation of emergency cycle lanes to support safe community during the pandemic. Despite these interventions, we are also seeing more commuting by private vehicles as people are trying to avoid public transport \citep{Forbes}, hence an understanding of their impact on cycle lanes on congestion is important. 

It is also worth noting that we are obliged to assume that the outcomes of one unit are not affected by the treatment assignment of any other units. While not always plausible, we attempt to reduce the “spillover” effects by reducing interactions between the treated and control units. One possible direction of further studies could be using an improved design and defining several types of treatment effects following \citet{Hudgens_2008} and \citet{Brian_2020}, and develop associated estimation methodologies for the setting where there may be clustered interference. Assessment of competing estimators of quantile effect at extreme tails which lies beyond the range of sample data could be another direction of future research. The scoring method of \cite{gandy2020scoring} for unconditional extreme quantile estimates can be extended to assess estimates of causal effects at extreme tails.

\section*{\small Acknowledgement}
The authors would like to acknowledge the Lloyd’s Register Foundation for funding this research through the programme on Data-Centric Engineering at the Alan Turing Institute.
The authors are grateful to Dr. Haojie Li for providing the London Traffic data set, and Dr. Almut Veraart and Dr. Joydeep Chowdhury for helpful comments and suggestions. The authors also like to express their sincere thanks to the two anonymous reviewers and the Associate Editor for many helpful comments which have improved the presentation of this paper.

\bibliographystyle{apalike}
\bibliography{Reference}

\begin{thebibliography}{}

\bibitem[Bader et~al., 2018]{bader2018}
Bader, B., Yan, J., and Zhang, X. (2018).
\newblock Automated threshold selection for extreme value analysis via ordered
  goodness-of-fit tests with adjustment for false discovery rate.
\newblock {\em Ann. Appl. Stat.}, 12(1):310--329.

\bibitem[Badoe and Miller, 2000]{Land1}
Badoe, D.~A. and Miller, E.~J. (2000).
\newblock Transportation-land-use interaction: empirical findings in north
  america, and their implications for modeling.
\newblock {\em Transportation Research Part D: Transport and Environment},
  5(4):235--263.

\bibitem[Balkema and de~Haan, 1974]{BalkemadeHaan1974GPD}
Balkema, A.~A. and de~Haan, L. (1974).
\newblock {Residual Life Time at Great Age}.
\newblock {\em The Annals of Probability}, 2(5):792 -- 804.

\bibitem[Barkley et~al., 2020]{Brian_2020}
Barkley, B.~G., Hudgens, M.~G., Clemens, J.~D., Ali, M., and Emch, E.~E.
  (2020).
\newblock Causal inference from observational studies with clustered
  interference, with application to a cholera vaccine study.
\newblock {\em Annals of Applied Statistics}, 14(3):1432--1448.

\bibitem[Beirlant et~al., 2004]{beirlant_2004}
Beirlant, J., Goegebeur, Y., Segers, J., Teugels, J., De~Waal, D., and Ferro,
  C. (2004).
\newblock {\em Statistics of Extremes: Theory and Applications}.
\newblock Wiley Series in Probability and Statistics. John Wiley \& Sons.

\bibitem[Beirlant et~al., 2005]{Beirlant2005}
Beirlant, J., Goegebeur, Y., Teugels, J., and Segers, J. (2005).
\newblock {\em Statistics of Extremes}.
\newblock Wiley-Blackwell, 1 edition.

\bibitem[Benjamini and Hochberg, 1995]{Benjamini1995}
Benjamini, Y. and Hochberg, Y. (1995).
\newblock Controlling the false discovery rate: A practical and powerful
  approach to multiple testing.
\newblock {\em Journal of the Royal Statistical Society. Series B
  (Methodological)}, 57(1):289--300.

\bibitem[Bhuyan et~al., 2021]{Bhuyan2021}
Bhuyan, P., McCoy, E.~J., Li, H., and Graham, D.~J. (2021).
\newblock Analysing the causal effect of london cycle superhighways on traffic
  congestion.
\newblock {\em The Annals of Applied Statistics}, 15(4):1999--2022.

\bibitem[Blunden, 2016]{Evening}
Blunden, M. (2016).
\newblock Cycle superhighways make traffic worse in the city, report reveals.
\newblock {\em EveningStandard}, Oct 5.

\bibitem[Bondell et~al., 2010]{Bondell_2010}
Bondell, H.~D., Reich, B.~J., and Wang, H. (2010).
\newblock Noncrossing quantile regression curve estimation.
\newblock {\em Biometrika}, 97(4):825--838.

\bibitem[Chen et~al., 2001]{Employ}
Chen, C., Jia, Z., and Varaiya, P. (2001).
\newblock Causes and cures of highway congestion.
\newblock {\em IEEE Control Systems Magazine}, 21(6):26--32.

\bibitem[Chernozhukov et~al., 2013]{victor2013}
Chernozhukov, V., Fernández-Val, I., and Melly, B. (2013).
\newblock Inference on counterfactual distributions.
\newblock {\em Econometrica}, 81(6):2205--2268.

\bibitem[Chernozhukov et~al., 2020]{victor2020}
Chernozhukov, V., Fernández-Val, I., and Melly, B. (2020).
\newblock Fast algorithms for the quantile regression process.
\newblock {\em Empirical Economics}, 62:7--33.

\bibitem[Coles, 2001]{Coles_2001}
Coles, S.~G. (2001).
\newblock {\em An Introduction to Statistical Modeling of Extreme Values}.
\newblock Springer, London.

\bibitem[Davison and Smith, 1990]{Davison1990}
Davison, A.~C. and Smith, R.~L. (1990).
\newblock Models for exceedances over high thresholds.
\newblock {\em Journal of the Royal Statistical Society. Series B
  (Methodological)}, 52(3):393--442.

\bibitem[Deuber et~al., 2021]{Deuber2022}
Deuber, D., Li, J., Engelke, S., and Maathuis, M.~H. (2021).
\newblock Estimation and inference of extremal quantile treatment effects for
  heavy-tailed distributions.
\newblock {\em arXiv:2110.06627}.

\bibitem[Drees et~al., 2000]{drees2000}
Drees, H., de~Haan, L., and Resnick, S. (2000).
\newblock How to make a hill plot.
\newblock {\em Ann. Statist.}, 28(1):254--274.

\bibitem[Díaz, 2017]{Diaz_2017}
Díaz, I. (2017).
\newblock Efficient estimation of quantiles in missing data models.
\newblock {\em Journal of Statistical Planning and Inference}, 190:39--51.

\bibitem[Embrechts et~al., 1997]{Embrechts1997}
Embrechts, P., Klüppelberg, C., and Mikosch, T. (1997).
\newblock {\em Modelling Extremal Events}.
\newblock Berlin: Springer.

\bibitem[Engelke, 2020]{Engelke2020causal}
Engelke, S. (2020).
\newblock Graphical models and causality for extreme events.

\bibitem[FarahaCarlos and Azevedo, 2017]{FarahaCarlos_2017}
FarahaCarlos, H. and Azevedo, L. (2017).
\newblock Safety analysis of passing maneuvers using extreme value theory.
\newblock {\em IATSS Research}, 41:12--21.

\bibitem[Firpo, 2007]{Firpo_2017}
Firpo, S. (2007).
\newblock Efficient semiparametric estimation of quantile treatment effects.
\newblock {\em Econometrica}, 75:259--276.

\bibitem[Fisher and Tippett, 1928]{fisher_tippett_1928}
Fisher, R.~A. and Tippett, L. H.~C. (1928).
\newblock Limiting forms of the frequency distribution of the largest or
  smallest member of a sample.
\newblock {\em Mathematical Proceedings of the Cambridge Philosophical
  Society}, 24(2):180--190.

\bibitem[Freedman, 2010]{David_2010}
Freedman, D.~A. (2010).
\newblock {\em Statistical Models and Causal Inference: A Dialogue with the
  Social Sciences}.
\newblock Cambridge University Press.

\bibitem[Frölich and Melly, 2013]{Frolich_2013}
Frölich, M. and Melly, B. (2013).
\newblock Unconditional quantile treatment effects under endogeneity.
\newblock {\em Journal of Business \& Economic Statistics}, 31:346--357.

\bibitem[Gandy et~al., 2021]{gandy2020scoring}
Gandy, A., Jana, K., and Veraart, A. E.~D. (2021).
\newblock Scoring predictions at extreme quantiles.
\newblock {\em Advances in Statistical Analysis}.

\bibitem[Gangl, 2010]{Gangl_2013}
Gangl, M. (2010).
\newblock Causal inference in sociological research.
\newblock {\em Annual Review of Sociology}, 36:21--47.

\bibitem[George, 1970]{Bus}
George, K.~A. (1970).
\newblock Transportation compatible land uses and bus-stop location.
\newblock {\em Transactions on The Built Environment}, 44.

\bibitem[Gissibl and Klüppelberg, 2018]{gissibl2018}
Gissibl, N. and Klüppelberg, C. (2018).
\newblock Max-linear models on directed acyclic graphs.
\newblock {\em Bernoulli}, 24(4A):2693--2720.

\bibitem[Gissibl et~al., 2017]{gissibl2017tail}
Gissibl, N., Klüppelberg, C., and Otto, M. (2017).
\newblock Tail dependence of recursive max-linear models with regularly varying
  noise variables.
\newblock {\em arXiv: 1701.07351}.

\bibitem[Gnecco et~al., 2019]{gnecco2019causal}
Gnecco, N., Meinshausen, N., Peters, J., and Engelke, S. (2019).
\newblock Causal discovery in heavy-tailed models.
\newblock {\em arXiv: 1908.05097}.

\bibitem[Gomes and Guillou, 2015]{Gomes2015}
Gomes, M.~I. and Guillou, A. (2015).
\newblock Extreme value theory and statistics of univariate extremes: A review.
\newblock {\em International Statistical Review}, 83(2):263--292.

\bibitem[G'Sell et~al., 2015]{GSell2015}
G'Sell, M.~G., Wager, S., Chouldechova, A., and Tibshirani, R. (2015).
\newblock Sequential selection procedures and false discovery rate control.
\newblock {\em Journal of the Royal Statistical Society: Series B (Statistical
  Methodology)}, 78(2):423--444.

\bibitem[Hall, 1990]{hall_1990}
Hall, P. (1990).
\newblock Using the bootstrap to estimate mean squared error and select
  smoothing parameter in nonparametric problems.
\newblock {\em Journal of Multivariate Analysis}, 32(2):177 -- 203.

\bibitem[Hannart and Naveau, 2018]{Hannart2018}
Hannart, A. and Naveau, P. (2018).
\newblock Probabilities of causation of climate changes.
\newblock {\em Journal of Climate}, 31(14):5507--5524.

\bibitem[Hosking and Wallis, 1987]{Hosking1987}
Hosking, J. R.~M. and Wallis, J.~R. (1987).
\newblock Parameter and quantile estimation for the generalized pareto
  distribution.
\newblock {\em Technometrics}, 29(3):339--349.

\bibitem[Hudgens and Halloran, 2008]{Hudgens_2008}
Hudgens, M.~G. and Halloran, M.~E. (2008).
\newblock Toward causal inference with interference.
\newblock {\em Journal of the American Statistical Association},
  103(482):832--842.

\bibitem[Imbens and Rubin, 2015]{Imbens_2015}
Imbens, G.~W. and Rubin, D.~B. (2015).
\newblock {\em Causal inference in statistics, social, and biomedical
  sciences}.
\newblock Cambridge University Press.

\bibitem[Jin and Rafferty, 2017]{Growth}
Jin, J. and Rafferty, P. (2017).
\newblock Does congestion negatively affect income growth and employment
  growth? empirical evidence from us metropolitan regions.
\newblock {\em Transport Policy}, 55:1--8.

\bibitem[Kang and Schafer, 2007]{DR}
Kang, J. D.~Y. and Schafer, J.~L. (2007).
\newblock Demystifying double robustness: A comparison of alternative
  strategies for estimating a population mean from incomplete data.
\newblock {\em Statistical}, 22(4):523–539.

\bibitem[Karmakar, 1984]{karmakar1984}
Karmakar, N. (1984).
\newblock A new polynomial-time algorithm for linear programming.
\newblock {\em Combinatorica}, 4:373–395.

\bibitem[Koenker, 2005]{koenker2005}
Koenker, R. (2005).
\newblock {\em Quantile regression}.
\newblock Cambridge University press.

\bibitem[Langousis et~al., 2016]{Langousis2016}
Langousis, A., Mamalakis, A., Puliga, M., and Deidda, R. (2016).
\newblock Threshold detection for the generalized pareto distribution: Review
  of representative methods and application to the noaa ncdc daily rainfall
  database.
\newblock {\em Water Resources Research}, 52(4):2659--2681.

\bibitem[Li et~al., 2017]{Li_2017}
Li, H., J., G.~D., and Liu, P. (2017).
\newblock Safety effects of the london cycle superhighways on cycle collisions.
\newblock {\em Accident Analysis and Prevention}, 99:90--101.

\bibitem[Litvinova and Silvapulle, 2020]{Litvinova_2020}
Litvinova, S. and Silvapulle, M.~J. (2020).
\newblock Consistency of full-sample bootstrap for estimating high-quantile,
  tail probability, and tail index.
\newblock {\em arXiv:2004.12639}.

\bibitem[Liu and Wu., 2011]{Liu_2010}
Liu, Y. and Wu., Y. (2011).
\newblock Simultaneous multiple non-crossing quantile regression estimation
  using kernel constraints.
\newblock {\em Journal of nonparametric statistics}, 23(2):415--437.

\bibitem[Melly, 2006]{Melly_2006}
Melly, B. (2006).
\newblock Estimation of counterfactual distributions using quantile regression.
\newblock {\em Rev. Labor Econ.}, 68:543–572.

\bibitem[Mhalla et~al., 2019]{mhalla2019}
Mhalla, L., Chavez-Demoulin, V., and Dupuis, D.~J. (2019).
\newblock Causal mechanism of extreme river discharges in the upper danube
  basin network.

\bibitem[Moodie et~al., 2018]{Erica}
Moodie, E. E.~M., Saarela, O., and Stephens, D.~A. (2018).
\newblock A doubly robust weighting estimator of the average treatment effect
  on the treated.
\newblock {\em Stat}, 7(1).

\bibitem[Moodie and Stephens, 2022]{Erica_2022}
Moodie, E. E.~M. and Stephens, D.~A. (2022).
\newblock Causal inference: critical developments, past and future.
\newblock {\em arXiv:2204.02231}.

\bibitem[Naveau et~al., 2020]{Naveau2020review}
Naveau, P., Hannart, A., and Ribes, A. (2020).
\newblock Statistical methods for extreme event attribution in climate science.
\newblock {\em Annual Review of Statistics and Its Application}, 7(1):null.

\bibitem[Norman, 2017]{Guardian}
Norman, W. (2017).
\newblock Bike lanes don't clog up our roads, they keep london moving.
\newblock {\em The Gaurdian}, Dec 1.

\bibitem[Pickands, 1975]{Pickands_1975}
Pickands, J. (1975).
\newblock Statistical inference using extreme order statistics.
\newblock {\em Ann. Statist.}, 3(1):119--131.

\bibitem[Reid, 2021]{Forbes}
Reid, C. (2021).
\newblock None of top five congested roads in u.k. feature adjacent cycleways.
\newblock {\em Forbes}, Dec 7.

\bibitem[Retallack and Ostendorf, 2019]{Retallack}
Retallack, A. and Ostendorf, B. (2019).
\newblock Current understanding of the effects of congestion on traffic
  accidents.
\newblock {\em Int J Environ Res Public Health}, 13(16(18)).

\bibitem[Ribes et~al., 2020]{Ribes2020}
Ribes, A., Thao, S., and Cattiaux, J. (2020).
\newblock Describing the relationship between a weather event and climate
  change: a new statistical approach.
\newblock {\em Journal of Climate}, 0(0):null.

\bibitem[Rosenbaum and Rubin, 1983]{Rosenbaum1983}
Rosenbaum, P. and Rubin, D.~B. (1983).
\newblock The central role of the propensity score in observational studies for
  causal effec.
\newblock {\em Biometrika}, 40:41--55.

\bibitem[Scarrott and MacDonald, 2012]{Scarrott_2012}
Scarrott, C. and MacDonald, A. (2012).
\newblock Review of extreme value threshold estimation and uncertainty
  quantification.
\newblock {\em REVSTAT}, 10(1).

\bibitem[Schnabel and Eilers, 2013]{Sabine_2013}
Schnabel, S.~K. and Eilers, P. H.~C. (2013).
\newblock Simultaneous estimation of quantile curves using quantile sheets.
\newblock {\em AStA Advances in Statistical Analysis}, 97:77--87.

\bibitem[Slawson, 2017]{Cost}
Slawson, N. (2017).
\newblock Traffic jams on major uk roads cost economy around £9bn.
\newblock {\em The Gaurdian}, Oct 18.

\bibitem[{Transport for London}, 2010]{Mayor}
{Transport for London} (2010).
\newblock Cycling revolution london.

\bibitem[{Transport for London}, 2011]{TFL11}
{Transport for London} (2011).
\newblock Barclays cycle superhighways evaluation of pilot routes 3 and 7.

\bibitem[{UN Environment Programme}, 2019]{UN}
{UN Environment Programme} (2019).
\newblock Cycling, the better mode of transport.
\newblock {\em
  \url{https://www.unep.org/news-and-stories/story/cycling-better-mode-transport}},
  June 11.

\bibitem[{United Nations}, 2021]{Cycle_day}
{United Nations} (2021).
\newblock World bicycle day.
\newblock {\em \url{https://www.un.org/en/observances/bicycle-day}}, June 03.

\bibitem[Xian et~al., 2021]{Xian_2021}
Xian, X., Ye, H., Wang, X., and Lie, K. (2021).
\newblock Spatiotemporal modeling and real-time prediction of
  origin-destination traffic demand.
\newblock {\em 63}, 1(77-89).

\bibitem[Xu et~al., 2018]{Xu_2018}
Xu, D., Daniels, M.~J., and Winterstein, A.~G. (2018).
\newblock A bayesian nonparametric approach to causal inference on quantiles.
\newblock {\em Biometrics}, 74:259--276.

\bibitem[Xu and Nusholtz, 2017]{Xu_2017}
Xu, L. and Nusholtz, G. (2017).
\newblock Application of extreme value theory to crash data analysis.
\newblock {\em The Stapp Car Crash Journal}, 61:287--298.

\bibitem[Zhang et~al., 2017]{Land2}
Zhang, K., Sun, D.~J., Shen, S., and Zhu, Y. (2017).
\newblock Analyzing spatiotemporal congestion pattern on urban roads based on
  taxi gps data.
\newblock {\em Journal of Transport and Land Use}, 10(1):675--694.

\bibitem[Zhang et~al., 2021]{Zhang_2021}
Zhang, N., J., G.~D., Hörcher, D., and Bansal, P. (2021).
\newblock A causal inference approach to measure the vulnerability of urban
  metro systems.
\newblock {\em Transportation}.

\bibitem[Zhang, 2018]{YichongZhang}
Zhang, Y. (2018).
\newblock {Extremal quantile treatment effects}.
\newblock {\em The Annals of Statistics}, 46(6B):3707 -- 3740.

\bibitem[Zhang et~al., 2012]{zhang2012}
Zhang, Z., Chen, Z., Troendle, J.~F., and Zhang, J. (2012).
\newblock Causal inference on quantiles with an obstetric application.
\newblock {\em Biometrics}, 68(3):697--706.

\bibitem[Zheng and Sayed, 2019]{Zheng_2019}
Zheng, L. and Sayed, T. (2019).
\newblock Application of extreme value theory for before-after road safety
  analysis.
\newblock {\em Transportation Research Record: Journal of the Transportation
  Research Board}, 2673.

\bibitem[Zou and Yuan, 2008]{zhu-csda-2008}
Zou, H. and Yuan, M. (2008).
\newblock Regularized simultaneous model selection in multiple quantiles
  regression.
\newblock {\em Computational Statistics \& Data Analysis}, 52(12):5296 -- 5304.

\end{thebibliography}
\end{document}